\documentclass{aa}  
\usepackage{graphicx,natbib}
\usepackage{txfonts}
\usepackage{hyperref}

\begin{document} 

   \title{Chromospherically active stars: Chemical composition of photospheres in 20 RS~CVn stars}

   \author{  B.~Bale\inst{1}
   \and
   G.~Tautvai\v{s}ien{\. e}\inst{1}  
     \and
          R.~Minkevi\v{c}i\={u}t\.{e}\inst{1}
         \and
          A.~Drazdauskas\inst{1}
          \and
          \v{S}.~Mikolaitis\inst{1}
          \and
          E.~Stonkut\.{e}\inst{1}
          \and
          M.~Ambrosch\inst{1}
          }

   \institute{Institute of Theoretical Physics and Astronomy, Vilnius University, Saul\.etekio av. 3, 10257 Vilnius, Lithuania \\
   \email{barkha.bale@ff.vu.lt}
    }

   \date{Received 28 October 2024 / Accepted  18 February 2025}

  \abstract
   {Various element transport processes modify the photospheric chemical composition of low-mass stars during their evolution. The most prominent one is the first dredge-up that occurs at the beginning of the red giant branch. Then, various extra-mixing processes, such as those caused by thermohaline- and/or rotation-induced mixing, come into action. The extent of the influence of stellar magnetic activity on alterations in stellar chemical composition is among the least studied questions.} 
  {To investigate how magnetic activity influences mixing in the atmospheres of magnetically active stars, we carried out a detailed study of C, N, and up to ten other chemical element abundances, as well as carbon isotope ratios in a sample of RS\,CVn stars.}   
{Using a differential model atmosphere method, we analysed high-resolution spectra that had been observed with the VUES spectrograph on the 1.65~m telescope at the Moletai Astronomical Observatory of Vilnius University. Carbon abundances were derived using the spectral synthesis of the ${\rm C}_2$ band heads at 5135 and 5635.5~{\AA}. We analysed the wavelength intervals 6470--6490~{\AA} and 7980--8005~\AA\ with CN features to determine nitrogen abundances. Carbon isotope ratios were determined from the $^{13}$CN line at 8004.7~\AA.  Oxygen abundances were determined from the [O\,\textsc{i}] line at 6300~{\AA}. Abundances of other chemical elements were determined from equivalent widths or spectral syntheses of unblended spectral lines.}      
{We determined the main atmospheric parameters and abundances of up to 12 chemical elements for a sample of 20 RS~CVn giants that represented different evolutionary stages. We determined that *29\,Dra, *b01\,Cyg, and V*\,V834\,Her, which are in the evolutionary stage below the red giant branch luminosity bump, already show evidence of extra-mixing in their lowered carbon isotope ratios. }       
{We provide observational evidence that in low-mass chromospherically active RS\,CVn stars, due to their magnetic activity, extra-mixing processes may start acting below the luminosity bump of the red giant branch.}

\titlerunning{Chromospherically active stars: Chemical composition of photospheres}
\authorrunning{Bale et al.}

\maketitle
%

\section{Introduction}
The very first discussion of RS Canum Venaticorum (RS\,CVn) stars was started by \cite{Keller&Limber1951}. A systematic observation of these stars was initiated by \cite{Hall1976}, who proposed that a binary system could be classified as an RS\,CVn system if it meets the following criteria: (i) an orbital period between 1 and 14 days, (ii) strong emission in the Ca\,\textsc{ii} H and K lines, and (iii) a hot component is present with a spectral type of F or G and a luminosity class of IV or V. The general properties of RS\,CVn systems were also described by \cite{Montesinos1988}. Understanding the general properties is essential for delving into the more complex aspects of these binary systems. Among the various characteristics of RS\, CVn stars, their chemical composition emerges as a particularly intriguing area of study. The chemical abundance in these stars offers insight into their current state and serves as a window into their past evolutionary processes. Furthermore, the intense magnetic activity within RS\, CVn stars can lead to unique alterations in their chemical makeup. To understand the observed coronal abundance anomalies and genuine abundance variations, a thorough comparison with photospheric abundances is also needed (\citealt{Tautvaisiene2010a}). The chemical element abundances in the photospheres of RS\,CVn-type stars are distinctive, which suggests the influence of multiple physical processes linked to their activity \citep{Pallavicini1992, Tautvaisiene1992, Randich1993, Randich1994, Savanov&Berdyugina1994, Berdyugina1998, Berdyugina1999, Katz2003, Morel2003, Morel2004, Barisevicius2010, Barisevicius2011, Tautvaisiene2010b, Tautvaisiene2011, Pakhomov2015, Xing2021}.

Carbon, nitrogen, and oxygen are key elements that reflect chemical enrichment and influence the lifetimes of stars, their energy production, and their positions on the Hertzsprung-Russell diagram. Their abundances, shaped by different stages of stellar evolution, offer insight into galactic and stellar evolution \citep{Tautvaisiene2015}. Oxygen, the third most abundant element in the universe, serves as a fundamental tracer for the formation and evolution of planets, stars, and galaxies. Unlike carbon, the oxygen abundance in stellar atmospheres remains relatively constant over the course of a star's lifetime (e.g. \citealt{Stonkute2020}). Carbon and nitrogen abundances are key quantitative indicators of mixing processes in evolved stars. During the first dredge-up, the abundance of \textsuperscript{12}C decreases, while the abundances of \textsuperscript{13}C and \textsuperscript{14}N  increase \citep{Iben1965}. It was later observed that stellar atmospheric abundances are influenced not only by the first dredge-up but also by extra-mixing \citep[and references therein]{Day1973, Pagel1974,Tomkin1976, Lambert&Ries1981, Gilroy1989, Gilroy&Brown1991, Tautvaisiene2000, Tautvaisiene2001, Tautvaisiene2005, Mikolaitis2010, Mikolaitis2011a, Mikolaitis2011b, Mikolaitis2012}.  Extra-mixing becomes efficient immediately after the luminosity function bump on the red giant branch (RGB), where the hydrogen-burning shell crosses the chemical discontinuity created by the outward-moving convective envelope \citep{Charbonnel1994, Charbonnel1998}. These alterations vary with stellar mass, metallicity, and evolutionary stage \citep{Lambert&Ries1977, Boothroyd&Sackmann1999, Charbonnel&Zahn2007}.

 The precise mechanisms underlying deep stellar mixing have long been a subject of intense study, yet they remain elusive, with various theoretical models proposed to explain these complex processes. Rotational mixing has been prominently discussed in the literature, with significant contributions from \cite{Palacios2003}, \cite{Chaname2005}, and \cite{Denissenkov2006}. Concurrently, the role of magnetic fields has been evaluated, notably in studies by \cite{Busso2007}, \cite{Nordhaus2008}, and \cite{Palmerini2009}. The synergistic effects of rotation and magnetic fields on stellar mixing have also been explored, with \cite{Eggenberger2005} investigating their combined impact and \cite{Denissenkov2009} detailing how magnetic buoyancy may enhance mixing in the radiative zones of red giants. The role of internal gravity waves, as proposed by \cite{Zahn1997} and \cite{Denissenkov&Tout2000}, alongside thermohaline mixing -- a process extensively outlined by researchers such as \cite{Eggleton2006, Eggleton2008}, \cite{Charbonnel&Zahn2007}, and \cite{Cantiello&Langer2010} -- represent additional facets of these complex interactions. Further elucidation of these dynamics is provided by \cite{Charbonnel&Lagarde2010} and \cite{Lagarde2012}, who investigated the  interplay between thermohaline mixing and rotational mixing in stars. The \textsuperscript{12}C/\textsuperscript{13}C ratio is a key diagnostic of deep mixing due to its sensitivity to mixing processes and its relative insensitivity to stellar parameters. A detailed study of chemical composition, including abundances of carbon isotopes, nitrogen, and oxygen, in chromospherically active RS~CVn stars may bring more clarity in this field of research. 

 RS\,CVn stars are chromospherically active binary systems that often showcase extensive star spots and large chromospheric plages, emitting intense flares that are observable across all wavelengths \citep{Zhilyaev2024, Buccino&Mauas2009}. They exhibit strong Ca\,\textsc{ii} H and K emissions, indicative of their dynamic chromospheres \citep{Biermann&Hall1976, Buccino&Mauas2009}. The emissions from active regions in these chromospherically active stars appear to be numerous and relatively evenly distributed across the stellar surface \citep{Sciortino1994, Schmitt&Stern1994}. However, there is some evidence suggesting that they can occasionally correlate with spots, similar to what is observed on the Sun \citep{Baliunas&Dupree1982}. Additionally, the random distribution of numerous spots with finite lifetimes, influenced by differential rotation, can account for most of the photometric variations observed in these stars \citep{Eaton1996}. In eclipsing RS CVn-type systems, the presence of cool spots contributes to significant variability in their light curves outside of eclipses \citep{Berdyugina2005}. These effects are also noticeable in the cores of spectral lines that form in the chromosphere, such as Ca\,\textsc{ii} H \& K and H$_\alpha$, where the core emissions can be much stronger than those seen in the Sun \citep{Strassmeier2000}. 

In this work, we investigate 20 single-line RS~CVn stars (*29\,Dra, *eps\,Leo, *39\,Cet, HD\,145742, V*\,V403\,Aur, HD\,191588, V*\,V834\,Her, *62\,Ser, *37\,Com, V*\,KX\,Peg, *ksi\,Leo, HD\,200740, *mu.\,Leo, *11\,LMi, BD+15\,3367, V*\,BF\,Psc, *b01\,Cyg, HD\,221639, *3\,Cam, and V*\,CL\,Cam). Ten of them were observed twice with different spectral resolutions. In this paper, we determined the chemical abundances of C(C$_2$), N(CN), [O\,\textsc{i}], Mg\,\textsc{i}, Si\,\textsc{i}, Ca\,\textsc{i}, Sc\,\textsc{i}, Sc\,\textsc{ii}, Ti\,\textsc{i}, Ti\,\textsc{ii}, Cr\,\textsc{i}, Cr\,\textsc{ii}, Fe\,\textsc{i}, Fe\,\textsc{ii}, Co\,\textsc{i}, and Ni\,\textsc{i}. Some of the investigated stars have had subsets of chemical element abundances reported in the literature, such as in the works by \cite{Luck2015} and \cite{Boyarchuk2001}. \cite{Tautvaisiene2010b} and \cite{Tautvaisiene2011} studied the chemical composition of $\lambda$\,And and AY\,Cet. \cite{Barisevicius2010} and  \cite{Barisevicius2011} provided a detailed chemical composition of *29\,Dra and *33\,Psc. There is no chemical composition information available in the literature for our sample stars HD\,191588, BD+15\,3367, V*\,CL\,Cam, HD\,221639, V*\,V834\,Her, and *3\,Cam. For more details, we refer to the following section on results and discussion. Hence, there remains a need to study the chemical element abundances and mixing processes inside these chromospherically active RS\,CVn stars. 

The structure of this paper is as follows. Sect.~\ref{sample} describes the stellar sample selection, observations, and analysis methods that cover the determination of atmospheric parameters, elemental abundances, carbon isotope ratios, as well as stellar masses, kinematics, and age estimations. In Sect.~\ref{Resultsdis}, we present and analyse the results, and provide a detailed discussion of each star. This section includes an examination of the mixing processes in RS\,CVn stars, an analysis of their chemical abundances, and an exploration of the relationship between magnetic activity and the evolutionary progression of RS\,CVn systems. Section~\ref{conclusion} provides a summary and conclusions.

\begin{figure*}
    \centering
        \includegraphics[width=1.0\textwidth]{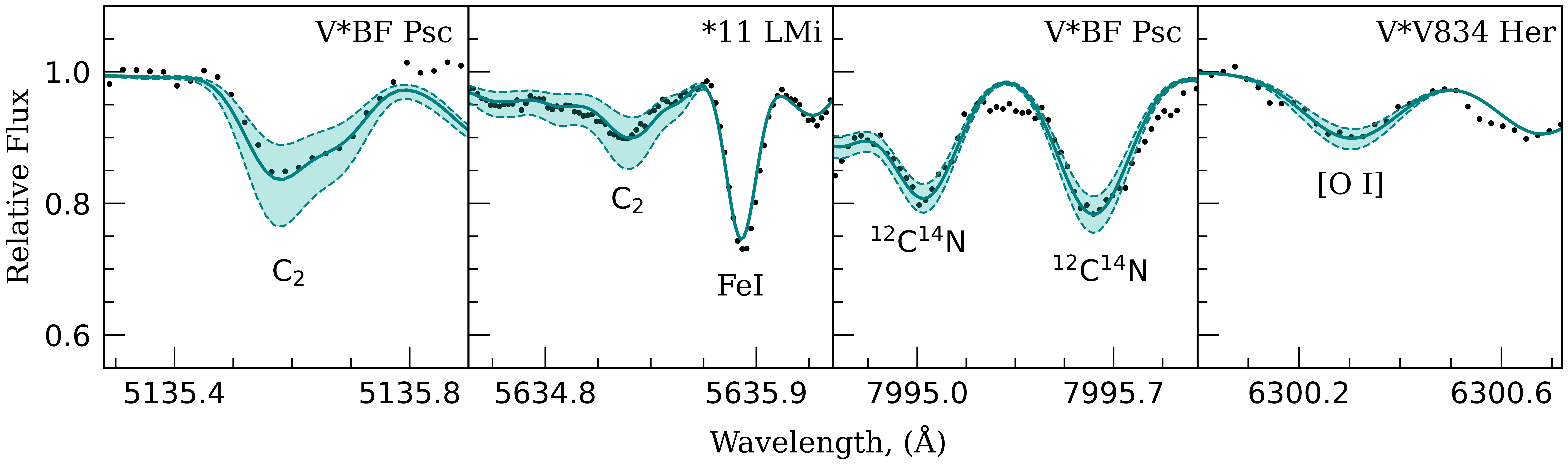}
            \includegraphics[width=1.0\textwidth]{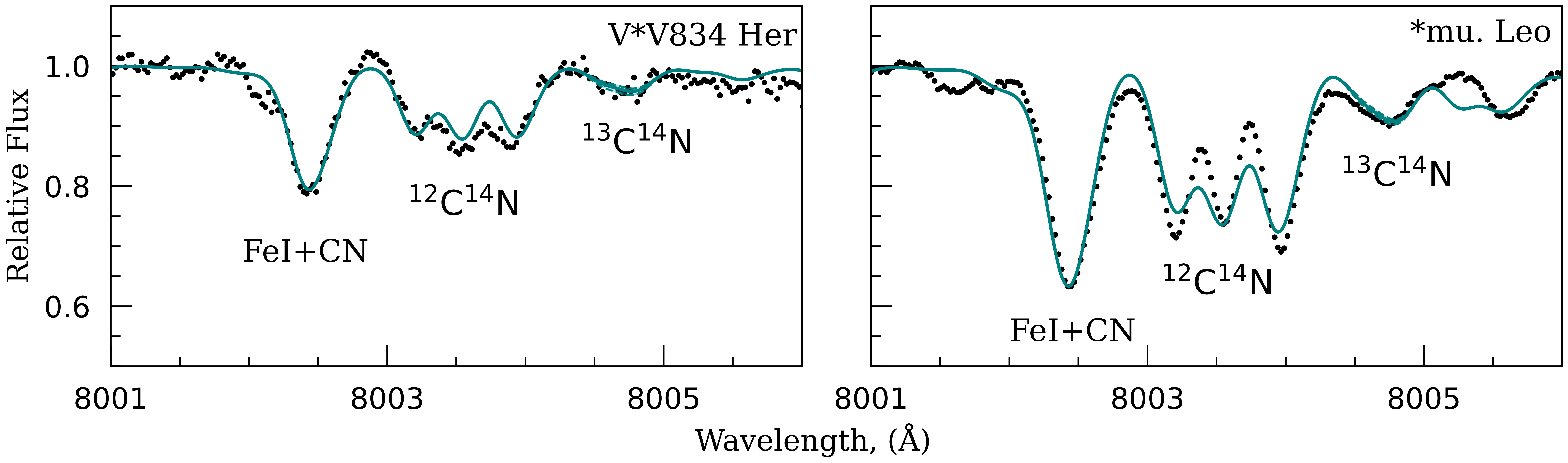}
    \caption{Examples of stellar spectra and spectral syntheses. The observed spectra are displayed using the black dots, while the solid green lines represent the optimal fit of synthetic spectra to the observed data, accounting for variations of $\pm 0.10$~dex in abundance marked by the filled region around the optimal fit.}
    \label{4o}
\end{figure*}

\section{Stellar sample  and analysis }
\label{sample}
\subsection{Sample selection}
\label{sample selection} 
The sample of 20 low-mass chromospherically active RS CVn stars for our study was selected from several sources in the literature. Specifically,  *eps\,Leo, V*\,V403\,Aur, V*\,V834 Her, *62\,Ser, *ksi\,Leo, HD\,200740, *mu.\,Leo, and *b01\,Cyg were selected from \cite{Strassmeier1994}, who investigated high-resolution spectra of evolved stars that exhibit chromospheric activity.  The stars HD\,191588, V*\,KX\,Peg, V*\,BF\,Psc, *3\,Cam, V*\,CL\,Cam, *29\,Dra, and *39\,Cet were selected  from \cite{Eker2008}, which provided a detailed description of emission features such as Ca\,\textsc{ii} H \& K and H$_\alpha$.  The stars HD\,145742, BD+15\,3367, and HD\,221639 were included based on references cited by \cite{Pace2013}.  The strength of the magnetic field of *37\,Com was determined by \cite{Auriere2015}. 
The selection of these 20 stars was based on several factors, including their brightness, positional coordinates, allocated observation time, and the resolution capabilities of the telescope.

\subsection{Observations}
\label{Observation} 
 
We observed the RS\,CVn stars using the 1.65-meter Ritchey-Chr{\' e}tien telescope located at the Moletai Astronomical Observatory, which is operated by Vilnius University in Lithuania. High-resolution spectra were obtained with the Vilnius University Echelle Spectrograph \citep[VUES;][]{Jurgenson2016}. VUES is capable of covering a spectral range from 4000 to 8800~\AA\ and offers three distinct resolution modes. For observations in this study, we used the spectral resolutions $R\sim 36\,000$ and $R\sim 68\,000$. 
The exposure times for the RS CVn stars were adjusted according to their brightness, ranging from 1200 to 3600~s. The resulting data had signal-to-noise ratios (S/N) between 75 and 200, with a median S/N of about 96.  All observations were carried out from 2021 to 2024. Data reductions were performed using the dedicated automated pipeline developed by \cite{Jurgenson2016}, which ensured efficient and consistent processing of the spectral data. This approach facilitated the detailed analysis of the spectra within the specified wavelength range and resolution. 

All 20 stars were initially observed at a resolution of $\sim 36\,000$. Ten of the stars were later observed with a resolution of $\sim68\,000$ as well. Information on the spectral resolution used is provided along with the results.

\subsection{Atmospheric parameters}
\label{Atmospheric parameters} 

To determine the stellar atmospheric parameters (effective temperature, $T_{\rm eff}$, surface gravity, log\,$g$, microturbulence velocity, $v_{\rm t}$, and metallicity [Fe/H]), we adopted the classical method of equivalent widths using atomic neutral and ionized iron lines. The measurement of equivalent widths was done using the Spectral Analysis Tool- Virtual Observatory (SPLAT-VO) software \citep{Skoda2014}. The equivalent widths of the iron lines were used to determine atmospheric parameters with the MOOG \citep{Sneden1973} code. This process was iterated until the determination of atmospheric parameters resulted in a zero trend in the relationships of [El/H] versus the excitation potential, and [El/H] versus the equivalent widths of  Fe\,\textsc{i} and Fe\,\textsc{ii} lines. On average 70 Fe\,\textsc{i} and 7 Fe\,\textsc{ii} lines were used for the analysis.  
The spectral analysis was based on plane-parallel, one-dimensional, hydrostatic model atmospheres under local thermodynamic equilibrium (LTE) conditions with constant flux, as provided by the Model Atmospheres with a Radiative and Convective Scheme (MARCS) stellar model atmosphere and flux library \citep{Gustafsson2008} as in the $Gaia$-ESO Survey, and the Solar abundances reported by \cite{Grevesse2007}. Atomic lines were selected from the $Gaia$-ESO line list by (\citealt{Heiter2015}).

\subsection{Elemental abundances and carbon isotope ratios}
\label{elemental abu& cc ratio para}

To derive the abundances of carbon, nitrogen, oxygen, and magnesium we applied the spectral synthesis method with the {TURBOSPECTRUM} code 
\citep{Plez2012}.
For the determination of oxygen abundance, we used the forbidden line at 6300.3~\AA. For the carbon abundance, we used two C\textsubscript{2} molecular bands at 5135~\AA\ and 5635~\AA. For the nitrogen abundance determination, we used $^{12}$C$^{14}$N molecular lines in the regions
6470--6490 and 7980--8005~\AA. The interval between 7990~\AA\ and 8010~\AA,\, which includes the \textsuperscript{13}C\textsuperscript{14}N bands, was used for the carbon isotope ratio analysis. 
We performed several iterative steps until the determinations of carbon and oxygen abundances converged, as abundances of these elements are interrelated due to the molecular equilibrium.  Following this, the derived values of carbon and oxygen were used to determine the nitrogen abundance and subsequently the carbon isotope ratios. We refer the reader to Fig.~\ref{4o} for an illustration of the spectral synthesis method applied to the fitting of prominent spectral features. Detailed methodologies and procedures can be found in the $Gaia$-ESO Survey paper by \cite{Tautvaisiene2015}. The impact of non-local thermodynamic equilibrium (NLTE) on C\textsubscript{2} Swan bands has not been extensively studied; however, it is believed that NLTE effects do not substantially influence the measurements of carbon abundance, as evidenced by consistent results from both the forbidden [C\,\textsc{i}] line and C\textsubscript{2} lines \citep{Gustafsson1999}. The forbidden oxygen line at 6300.3~\AA\ has been comprehensively studied, leading to the conclusion that LTE accurately models this line \citep{Asplund2005}. To determine the Mg\,\textsc{i}, abundance, we utilized four spectral lines: 5528.41, 5711.07, 6318.71, and 6319.24~\AA. 
For *3\,Cam and V*\,CL\,Cam the oxygen spectral lines were contaminated by telluric interference, which complicated the direct determination of the oxygen abundance. Therefore, magnesium abundances were accepted as an alternative indicator for the oxygen abundance in these stars (cf. \citealt{Pagel1995}). To determine the abundances of other chemical elements, we employed a method of equivalent widths. The equivalent widths were measured using DAOSPEC \citep{Stetson&Pancino2008}, and elemental abundances for Si, Ca, Sc\,\textsc{i}, Sc\,\textsc{ii}, Ti\,\textsc{i}, Ti\,\textsc{ii}, Cr\,\textsc{i}, Cr\,\textsc{i}, Co, and Ni were determined using the MOOG code \citep{Sneden1973}. The number of lines utilized for each element is specified alongside the corresponding abundance results.

\subsection{Stellar masses and ages}
\label{Stellar masses} 

To determine the ages of the stars in our study, we employed the UniDAM tool developed by \cite{Mints&Hekker2017, Mints&Hekker2018}, which integrates Bayesian techniques and utilizes PAdova and TRieste Stellar Evolution Code (PARSEC) isochrones \citep{Bressan2012}. Our analysis used stellar atmospheric parameters derived from our own measurements, in conjunction with $J$, $H$, and $K$ magnitudes from the Two Micron All-Sky Survey (2MASS, \citealt{Skrutskie2006}), as well as $W1$ and $W2$ magnitudes from the All Wide-field Infrared Survey Explorer (AllWISE) (\citealt{Cutri2014}). We conducted a cross-match with the WISE catalog for our sample of stars, using a search radius of five arcseconds, and subsequently refined our dataset by extracting relevant parameters required for the UniDAM input files. To determine stellar masses, we implemented the quality flags specified in Sect.~6.1 of \cite{Mints&Hekker2017, Mints&Hekker2018}, and examined the evolutionary stages including the red giant branch (pre-core-helium burning; stage I), core-helium burning stars (stage II), and the asymptotic giant branch (post-core-helium burning; stage III).

\subsection{Kinematics properties}
\label{Kinematics} 
\begin{figure}
    \includegraphics[width=\columnwidth]{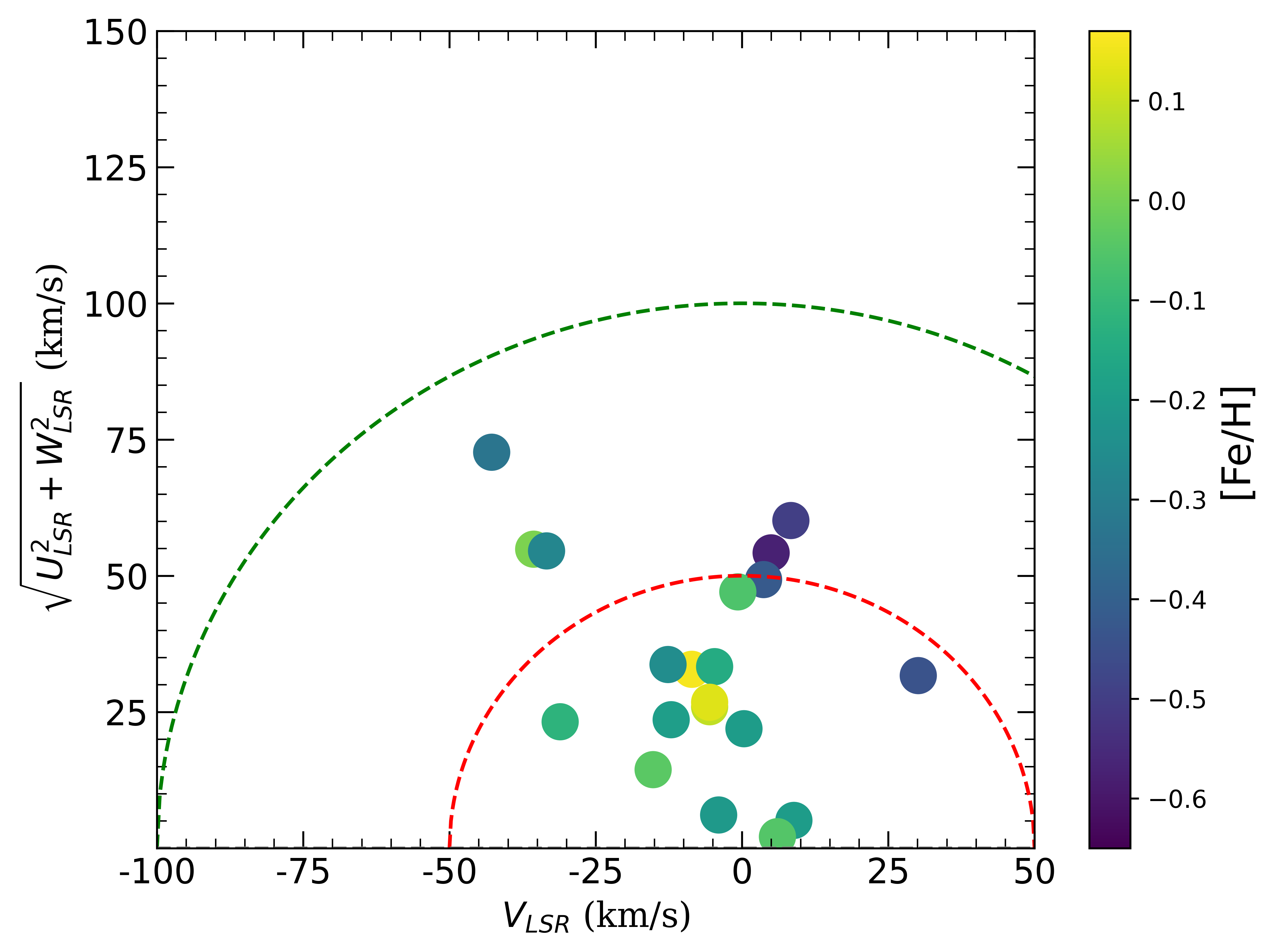}
    \caption{Toomre diagram for the stars in our sample. The dashed lines represent the constant total space velocity. The symbols have been colour-coded according to the stellar metallicity [Fe/H].}
    \label{uvw}
\end{figure}

\begin{figure}
    \includegraphics[width=\columnwidth]{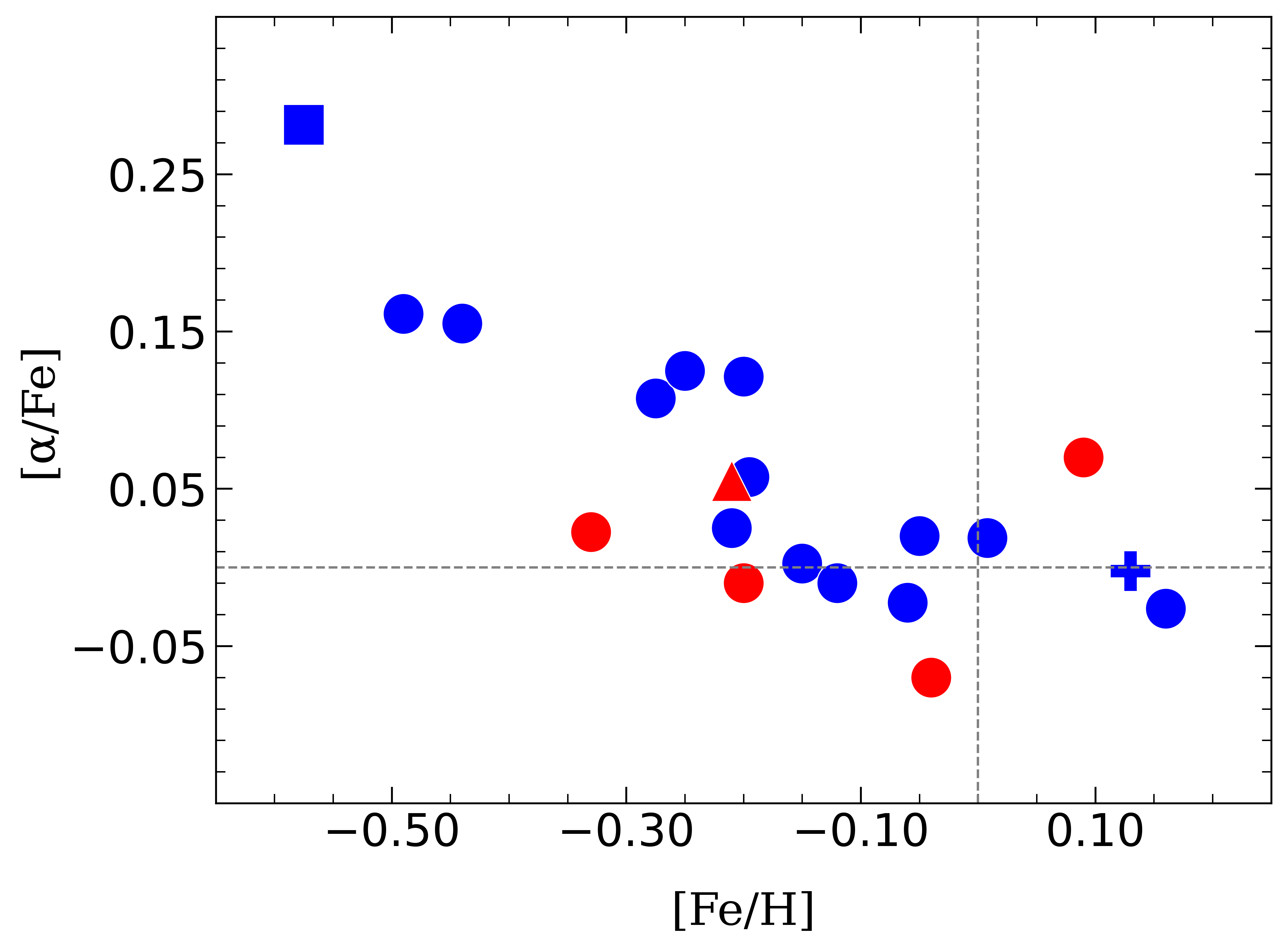}
    \caption{Observed $\alpha$-to-iron element abundance ratios as a function of [Fe/H]. The value of $[\alpha/\text{Fe\,\textsc{i}}]$ is computed as an average of the following elements: Mg\,\textsc{i}, Si\,\textsc{i}, Ca\,\textsc{i}, and Ti\,\textsc{i}. The meaning of symbols: the square is for the thick disc star, the circles are for the red giant branch stars, the plus sign marks a subgiant star, and the triangle is for the red clump star. The blue symbols mean that a star is at the evolutionary stage below the red giant luminosity bump, and the red is for stars at the evolutionary stage above the RGB luminosity bump.  
}
    \label{alpha}
\end{figure}

We computed the Galactic space velocities ($U$, $V$, $W$) and the mean galactocentric distances using the Galactic-dynamics calculation package $galpy$\footnote{http://github.com/jobovy/galpy} (\citealt{Bovy2015}) in Python. Observational data were sourced from multiple references for $galpy$ input: stellar distances from \cite{Bailer-Jones2021}, and other stellar parameters including proper motions and coordinates from the  \textit{Gaia} Data Release 3 \cite{GaiaCollaboration2021}. The radial velocity values were taken from our calculations. In Fig.~\ref{uvw}, we present the velocity components of our sample stars within a Toomre diagram, a tool for discerning the Galactic population structures -- namely, the thin disc, thick disc, and halo — based solely on stellar kinematics. Following criteria established  by \cite{Yoshii1982}, \cite{Gilmore&Reid1983}, and \cite{Recio-Blanco2014}, the  stars located outside the red circle marking the $\pm$50~km\,$s^{-1}$ velocity are considered as members of the thick disc.  Conversely, stars within this boundary are considered as part of the thin disc. However, after reviewing the Toomre diagram, atmospheric parameters, and the $[\alpha/\text{Fe}]$ versus [Fe/H] trend in Fig.~\ref{alpha}\footnote{In all plots we present the averaged values of parameters determined from spectra observed in the two spectral resolutions ($R=34\,500$ and $R=60\,000$), if available.}, we only accepted HD\,191588, the most metal deficient star with a relatively large [$\alpha$/Fe] value, as the thick disc star and the remaining stars are attributed to the thin disc in our study.

\subsection{Estimation of uncertainties }
\label{uncertainties} 
In this study, we encountered various potential sources of uncertainties and assessed them at each stage of the analysis. Systematic and random uncertainties must be taken into account. Systematic uncertainties can arise from inaccuracies in atomic data; however, these were largely mitigated by using differential analysis relative to the Sun. Random uncertainties can arise from the placement of the local continuum and the fitting of specific spectral lines.

In order to mitigate uncertainties in the determination of the main atmospheric parameters, we used a carefully selected set of Fe\,\textsc{i} and Fe\,\textsc{ii} lines. These lines were specifically chosen to avoid contamination from blends, telluric lines, or areas with challenging continuum determination.  The mean values of uncertainties in the full stellar sample are \( \sigma_{T_{\text{eff}}} = 50 \, \text{K} \),  \( \sigma_{\log g} = 0.18 \,  \text{dex} \),  \( \sigma_{[\text{Fe/H}]} = 0.08 \,  \text{dex} \), and  \( \sigma_{v_t} = 0.22 \,  \text{km s}^{-1} \). 

Uncertainties in elemental abundances can result from variations in atmospheric parameters. These uncertainties can be quantified by assessing the impact of deviations in each individual atmospheric parameter on the calculated abundances while keeping other parameters fixed. In Table~\ref{tab:errtab}, we provide the uncertainty values computed for the star *29\,Dra. 

\begin{table}
\centering
\caption{Effect of uncertainties in atmospheric parameters on the derived chemical abundances for the star *29\,Dra.}
\label{tab:errtab}
\begin{tabular}{lccccccl}
\hline
\hline
El.     & $\Delta T_{\rm eff}$  & $\Delta {\rm log}\,g$  & $\Delta$[Fe/H]  & $\Delta v_{\rm t}$ \\
        &  $\pm$50 K  & $\pm$0.18 dex    & $\pm$0.08 dex     &$\pm 0.22~{\rm km\, s}^{-1}$ \\
\hline
\hline
C(C\textsubscript{2})     & $\pm$0.01   & $\pm$0.01    & $\mp$0.02  & $\mp$0.02 \\
N(CN)  & $\pm$0.03   & $\pm$0.04  & $\pm$0.04  & $\mp$0.00  \\
O([O\,\textsc{i}])     & $\pm$0.01   & $\pm$0.08 & $\mp$0.01  & $\pm$0.00 \\ 
Mg\,\textsc{i}   & $\pm$0.03   & $\mp$0.02  & $\mp$0.02  & $\mp$0.08 \\ 
Si\,\textsc{i}     & $\pm$0.02   & $\pm$0.04  & $\pm$0.01  & $\pm$0.03 \\ 
Ca\,\textsc{i}   & $\pm$0.05   & $\mp$0.03  & $\pm$0.01  & $\mp$0.09 \\ 
Sc\,\textsc{i}     & $\pm$0.07   & $\pm$0.00  & $\pm$0.00  & $\mp$0.03 \\ 
Sc\,\textsc{ii}     & $\mp$0.00   & $\pm$0.08  & $\pm$0.02  & $\mp$0.08 \\ 
Ti\,\textsc{i}     & $\pm$0.07   & $\pm$0.00  & $\mp$0.00  & $\mp$0.07 \\ 
Ti\,\textsc{ii}     & $\mp$0.01   & $\pm$0.08  & $\pm$0.02  & $\mp$0.07 \\ 
Cr\,\textsc{i}     & $\pm$0.05   & $\mp$0.01  & $\pm$0.00  & $\mp$0.06 \\ 
Cr\,\textsc{ii}     & $\mp$0.04   & $\pm$0.08  & $\pm$0.01  & $\mp$0.04 \\ 
Co\,\textsc{i}    & $\pm$0.03   & $\pm$0.04  & $\pm$0.01  & $\mp$0.04 \\ 
Ni\,\textsc{i} & $\mp$0.02   & $\pm$0.04  & $\pm$0.01  & $\mp$0.09 \\ 
\hline
\end{tabular}
\end{table}

The random uncertainties can be evaluated from the scatter of abundances determined from various lines. The standard deviation for the oxygen abundance cannot be determined as we investigated only one line. The most probable uncertainty on oxygen abundance determinations is  \( \sigma_{[\text{O/Fe}]} \sim 0.05 \)~dex, since for the carbon abundances determined from two lines, the averaged scatter from all stars is only \( \sigma_{[\text{C/Fe}]} = 0.01 \)~dex, and from ten nitrogen lines is \( \sigma_{[\text{N/Fe}]} = 0.08 \)~dex.  For the \( ^{12}\text{C}/^{13}\text{C} \) ratio, the uncertainty depends on the strength of the $^{13}$CN feature, but the mean value is about $\pm 2$. 

Given the complex interdependencies among carbon, nitrogen, and oxygen abundances resulting from molecular equilibrium in stellar atmospheres, we examined how errors in one element could influence the measurements of the others. Our investigation reveals that an increase of $\Delta$[O/H] = 0.10 leads to changes of $\Delta$[C/H] = 0.03 and $\Delta$[N/H] = 0.07. Likewise, a modification of $\Delta$[C/H] = 0.10 causes  $\Delta$[N/H] = 0.10 and $\Delta$[O/H]  = 0.01. On the other hand, a change of $\Delta$[N/H]= 0.10 causes $\Delta$[C/H]= 0.00 and $\Delta$[O/H]= 0.01. 

The mean standard deviation of magnesium abundances calculated from the scatter in four lines is \( \sigma_{[\text{Mg\,\textsc{i},/Fe}]} = 0.07 \)~dex. There are very similar random uncertainties for other chemical elements as well. They vary from $\pm 0.06$~dex for [Cr\,\textsc{i}/H] to $\mp 0.11$~dex for [Ti\,\textsc{ii}/H]. 

The uncertainties in the determination of mass and age were calculated using the UniDAM (\citealt{Mints&Hekker2017, Mints&Hekker2018}) tool. The individual values of uncertainties are provided along with the mass and age results in machine-readable Table ~\ref{table:Results}.

\section{Results and discussion}
\label{Resultsdis}

\subsection{Stellar parameters}
\label{resultstellar} 

\begin{table*}
\centering
\caption{Main atmospheric parameters.}
\label{tab:stellar_parameters}
\begin{tabular}{lccccrcrcccc}
\hline
\hline
Star      & $T_{\rm eff}$ & $e_{T_{\rm eff}}$ & log\,$g$ & $e_{{\rm log}g}$ & [Fe/H] & $e_{\rm Fe\,I}$ & n$_{\rm Fe\,I}$ & $e_{\rm Fe\,II}$ & n$_{\rm Fe\,II}$ & $V_{\rm t}$ & $e_{V_{\rm t}}$ \\ 
    &  K & K &   &  &  &  &  &  &  & km\,s$^{-1}$  &  km\,s$^{-1}$ \\
\hline
*39 Cet   & 4940 & 32      & 2.81  & 0.13    & $-0.46$ & 0.06  & 74   & 0.06    & 6  & 1.37  & 0.14  \\ 
*39 Cet$^{\dagger}$ & 4920 & 41    & 2.87  & 0.19    & $-0.51$ & 0.09   & 97    & 0.11    & 15 & 1.76  & 0.20   \\ 
V* BF Psc    & 4990 & 68      & 3.37  & 0.21    & $-0.26$ & 0.10    & 65  & 0.02     & 5  & 0.78 & 0.35  \\ 
V* BF Psc$^{\dagger}$  & 4896 & 65      & 3.34  & 0.18    & $-0.29$ & 0.09   & 61   & 0.10    & 10 & 0.72  & 0.34  \\ 
*3 Cam       & 4705 & 55       & 2.27 & 0.19   & $-0.15$ & 0.10    & 67 & $-0.15$ & 7 & 1.29 & 0.19 \\
V* CL Cam    & 4700 & 50      & 2.61 & 0.19   & $-0.25$ & 0.11     & 72 & $-0.25$ & 9 & 1.47 & 0.14 \\
V* V403 Aur & 4870 & 42      & 2.32  & 0.19    & $-0.20$  & 0.09  & 68  & 0.10      & 5  & 1.47  & 0.18  \\ 
*ksi Leo      & 4730 & 62      & 2.54  & 0.17    & $-0.12$ & 0.09   & 65  & 0.06    & 6  & 1.41 & 0.20   \\ 
*11 LMi       & 5450 & 32      & 4.35  & 0.08    & 0.13  & 0.03   & 52  & 0.05    & 6  & 1.17 & 0.13  \\ 
*eps Leo  & 5440 & 36    & 2.31  & 0.15    & $-0.05$ & 0.07  & 74   & 0.05    & 5  & 1.83  & 0.17  \\ 
*mu. Leo       & 4470 & 75      & 2.41  & 0.14    & 0.09  & 0.08   & 55  & 0.04    & 7  & 1.72 & 0.23  \\ 
*37 Com       & 4535 & 68      & 1.90   & 0.20     & $-0.21$ & 0.11   & 66  & 0.13     & 10 & 1.80   & 0.21  \\ 
HD 145742 & 4730 & 38      & 2.53  & 0.18    & $-0.04$ & 0.09  & 75  & 0.07     & 5  & 1.20   & 0.22  \\ 
HD 145742$^{\dagger}$  & 4790 & 40    & 2.73 & 0.14     &$0.01$ & 0.08  & 71  & 0.08     &5   & 
1.20    &0.16 \\
*29 Dra   & 4745 & 40      & 2.87  & 0.18    & $-0.17$ & 0.10  & 77  & 0.06    & 9  & 1.64  & 0.23  \\ 
*29 Dra$^{\dagger}$   & 4738 & 62      & 2.79  & 0.18    & $-0.22$ & 0.09   & 63   & 0.08    & 6  & 1.65 & 0.23  \\ 
V* V834 Her & 4985 & 47      & 2.74  & 0.17    & $-0.06$ & 0.08  & 70   & 0.03    & 5  & 1.07  & 0.22  \\ 
BD+15 3367   & 4380 & 68      & 2.17  & 0.22    & $-0.33$ & 0.13   & 71  & 0.13     & 10 & 0.87  & 0.29  \\ 
*62 Ser      & 4660 & 52      & 2.54  & 0.16    & $-0.19$ & 0.09  & 72  & 0.07    & 5  & 1.09  & 0.20   \\ 
*62 Ser$^{\dagger}$      & 4690 & 50      & 2.49  & 0.20     & $-0.23$ & 0.11   & 112  & 0.10    & 16 & 1.23  & 0.18  \\ 
*b01 Cyg      & 5085 & 37      & 3.67  & 0.12    & 0.02  & 0.06  & 73  & 0.04    & 6  & 0.74  & 0.25  \\ 
*b01 Cyg$^{\dagger}$     & 5116 & 45      & 3.70   & 0.13    & $-0.01$ & 0.06  & 65  & 0.04     & 6  & 0.77 & 0.24  \\ 
HD 191588   & 4590 & 70      & 2.69  & 0.19    & $-0.55$ & 0.10  & 52   & 0.09    & 6  & 1.49  & 0.25  \\ 
HD 191588$^{\dagger}$   & 4547 & 65      & 2.56  & 0.17    & $-0.60$  & 0.09   & 44   & 0.08     & 6  & 1.54 & 0.24  \\ 
HD 200740    & 4920 & 50      & 2.87  & 0.16    & 0.17  & 0.08  & 70  & 0.07    & 7  & 1.10   & 0.23  \\ 
HD 200740$^{\dagger}$   & 4966 & 55      & 2.96  & 0.08    & 0.15  & 0.09   & 69  & 0.08    & 7  & 1.22 & 0.21  \\ 
V* KX Peg    & 5050 & 60      & 3.25  & 0.21    & $-0.46$ & 0.09   & 62  & 0.08    & 7  & 1.08  & 0.25  \\ 
V* KX Peg$^{\dagger}$   & 5072 & 60      & 3.38  & 0.19    & $-0.42$ & 0.09   & 54  & 0.09    & 7  & 1.08 & 0.24  \\ 
HD 221639   & 4960 & 45      & 3.08  & 0.19    & $-0.19$ & 0.09   & 80   & 0.06     & 8  & 0.92 & 0.24 \\
HD 221639$^{\dagger}$  &4965 & 40      & 2.86  & 0.16    & $-0.21$  & 0.08   & 75   & 0.06     & 7  & 0.96 &  0.19 \\
\hline
\end{tabular}
\tablefoot{$^{\dagger}$ Results obtained from observations with $R\sim68\,000$. Other results were obtained with $R\sim36\,000$.}
\end{table*}

We aimed to investigate stars at different post-main-sequence evolutionary stages. 
The determined atmospheric parameters are presented in Table~\ref{tab:stellar_parameters} for observations in both resolutions.  
The effective temperature, $T_{\rm eff}$, of the observed stars ranges from 4380 to 5450~K, with a mean temperature of 4855~K. 

The surface gravity in our stellar sample ranges from 1.9 to 4.35~dex, with a mean value of 2.83~dex. 
The lower log\,$g$ values indicate giant stars with expanded and less dense atmospheres. In comparison, a star with a higher value of 4.35~dex may be a subgiant transitioning from the main sequence.
Metallicity ([Fe/H]) spans from $-0.6$ to 0.17~dex, with a mean value of $-0.19$~dex, suggesting a slight metal deficiency in the sample.
The microturbulent velocity, $v_{\rm t}$, in the investigated stars ranges from 0.72 to 1.83~km\,$s^{-1}$, with a mean of 1.25~km\,$s^{-1}$. Microturbulent velocity helps to evaluate line broadening caused by turbulence in stellar atmospheres.

The determined ages span quite a large interval from 224~Myr to 11.7~Gyr, and masses from  3.63~$M_\odot$ to 0.86~$M_\odot$.  
The determined parameters indicate that the RS\,~CVn stars in this study represent a diverse group, which primarily encompasses giants at different evolutionary stages. 

\subsection{Comparison with other studies}
\label{comparison}

\begin{figure*}
    \centering
    \includegraphics[width=0.34\textwidth]{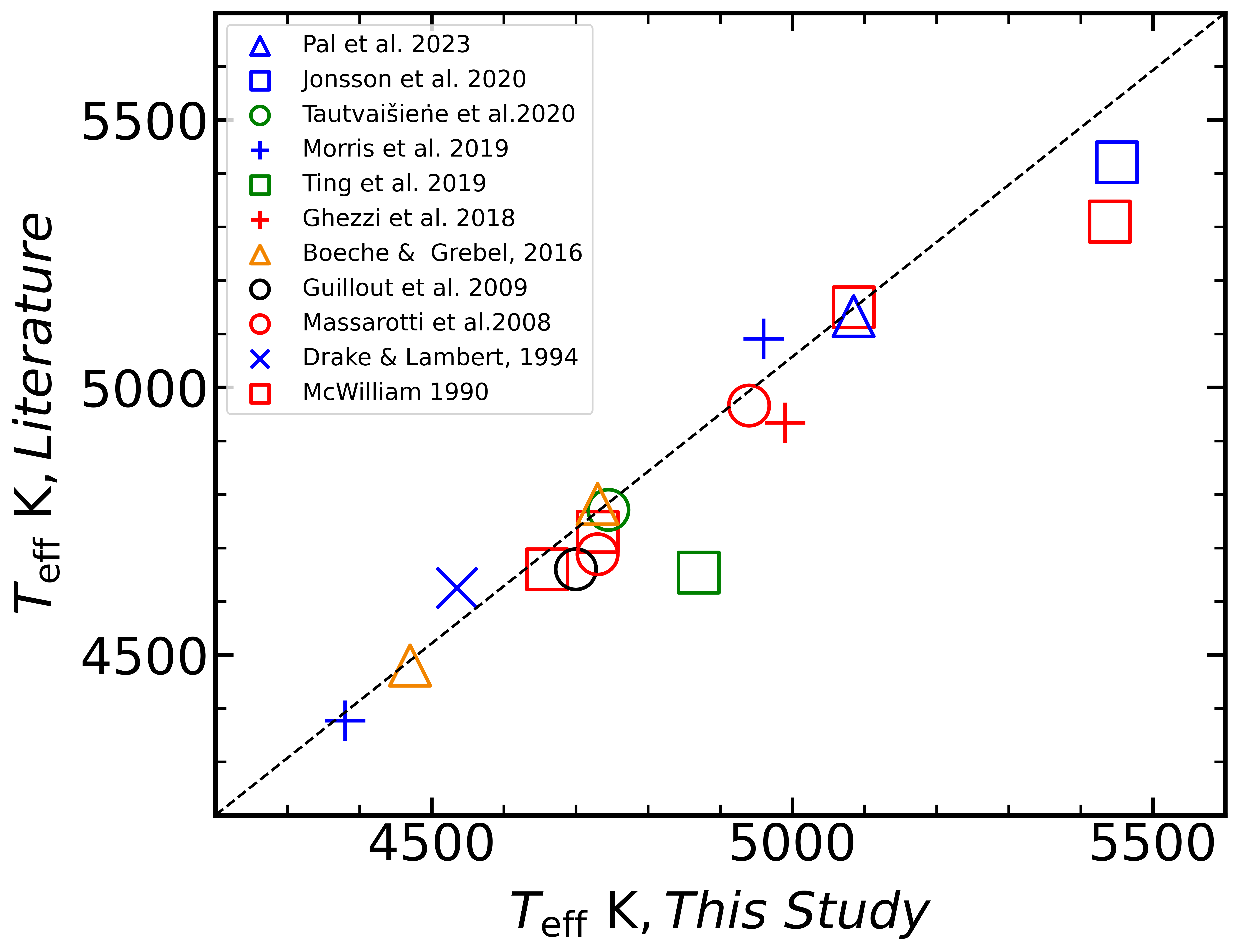}
    \includegraphics[width=0.31\textwidth]{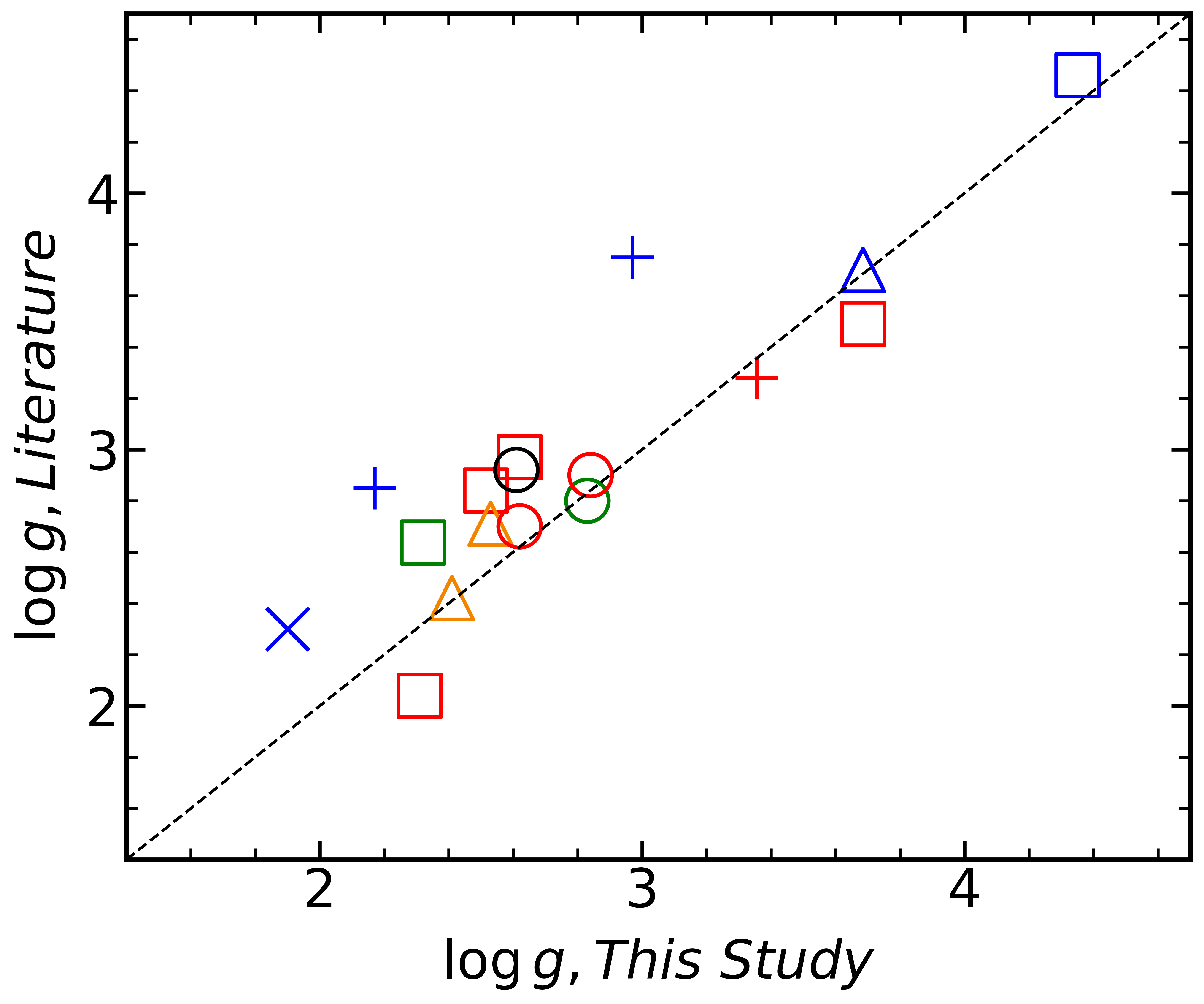}
    \includegraphics[width=0.34\textwidth]{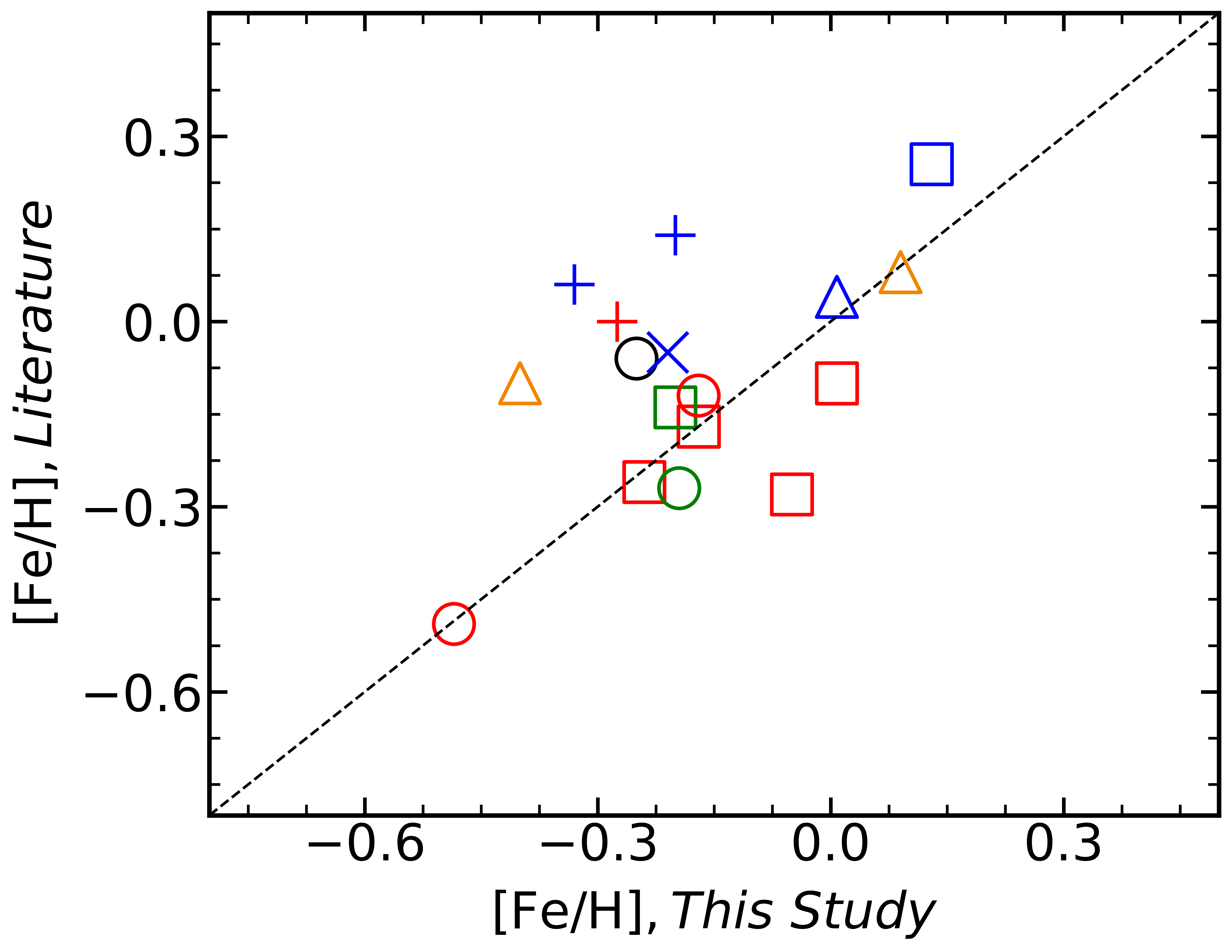}
    \caption{Comparison of the atmospheric parameters,  $T_{\rm eff}$, log\,$g$,  and [Fe/H] determined in our study  with results from studies listed in Sect.~\ref{comparison}.}
    \label{comppp}
\end{figure*}

We compared our atmospheric parameters, $T_{\rm eff}$, log\,$g$, and [Fe/H] with existing data from the literature (Fig.~\ref{comppp}). The comparison is made with spectroscopic studies for four stars (*b01\,Cyg, *ksi\,Leo, *62\,Ser, *eps\,Leo) in common with \cite{McWilliam1990}, and with one or two stars in common with other recent studies which include:  
\cite{Tautvaisiene2020} for *29\,Dra;
\cite{Jonsson2020} for *11\,LMi;
\cite{Morris2019} for BD+15\,3367 and HD\,221639;
\cite{Ting2019} for V*\,V403\,Aur;
\cite{Ghezzi2018} for V*\,BF\,Psc;
\cite{Boeche&Grebel2016} for *mu.\,Leo, HD\,145742; and
\cite{Guillout2009} for V*\,CL\,Cam. 
A comparison was also made with the photometric determinations by 
\cite{Pal2023} for *b01\,Cyg; 
\cite{Massarotti2008} for *39\,Cet and *ksi\,Leo; and
\cite{Drake1994} for *37\,Com.

 In the context of the comparison temperature and surface gravity, the results of our study align closely with the values reported in other studies. The determinations of metallicity exhibit some scatter, and the discrepancies reach $\pm 0.3$~dex in several cases, which underscores a general agreement in the metallicity estimates among the different studies.

\subsection{Ratios of \textsuperscript{12}C/\textsuperscript{13}C and  C/N } 
\label{cn} 

The carbon and nitrogen abundance ratio, C/N, and the carbon isotope ratio, \textsuperscript{12}C/\textsuperscript{13}C, are key indicators for understanding stellar evolution. The abundance alterations in these ratios that manifest in atmospheres of RS\, CVn stars offer valuable insight into stellar evolutionary processes.

In the previous high-resolution spectroscopic studies of RS\,CVn stars accomplished by \cite{Tautvaisiene2010b} and \cite{Drake2011} for $\lambda$\,And, and by \cite{Barisevicius2010} for 29\,Dra, it was inferred that extra-mixing processes may start acting
in low-mass highly chromospherically active stars below the bump of the luminosity function of the RGB. For the less active stars, this effect was not detected \citep{Barisevicius2011, Tautvaisiene2011}.

In terms of stellar structure, the RGB bump corresponds to the moment when the hydrogen-burning shell encounters the chemical
discontinuity created by the convective envelope at its maximum
penetration during the first dredge-up. When the hydrogen-burning shell reaches the H-rich previously mixed zone, the corresponding
decrease in molecular weight of the H-burning layers induces
a temporary drop in total stellar luminosity, which creates a bump in the luminosity function. During this stellar evolutionary stage the so-called extra mixing starts acting and lasts at least until the tip of the RGB. Among the most probable triggers of this extra-mixing is thermohaline-induced mixing (\citealt{Charbonnel&Zahn2007, Lagarde2019}, and references therein). For stars with masses above $\sim 2.2\,M\odot$ , no thermphaline-induced extra mixing is foreseen on the RGB (cf. \citealt{Lagarde2019}) and another so-called second dredge-up only starts on the asymptotic giant branch (e.g. \citealt{Boothroyd&Sackmann1999}), unless rotation-induced mixing emerges (\citealt{Charbonnel&Lagarde2010}). During the RGB extra-mixing phase, similarly to during the first dredge-up, the chemical elements $^{13}$C and $^{14}$N again diffuse outwards, while $^{12}$C diffuses inwards, which implies an additional decrease in the  \textsuperscript{12}C/\textsuperscript{13}C and  C/N ratios (\citealt{Charbonnel&Lagarde2010, Lagarde2017}, and references therein). For the \textsuperscript{12}C/\textsuperscript{13}C ratios, these abundance alterations are much more pronounced than for C/N ratios and are more useful for the investigation of extra mixing.  

Thus in our study, it was important to know if a star is below or above the RGB luminosity bump in its evolution. We carefully investigated the positions of stars in the log\,$g$ versus $T_{\rm eff}$ diagram using the PADOVA evolutionary tracks taken from \cite{Bressan2012}. The RGB luminosity bump is visible in the stellar evolutionary sequences as a zigzag. For example, for a 1\,$M_\odot$ solar-metallicity star, the luminosity bump is at about log\,$g \sim 2.55$ and $T_{\rm eff}\sim 4550$~K.

In our sample of 20 stars, 15 stars are below the RGB luminosity bump (we mark them in our work as BB) and five above (marked as AB). One of the BB stars (*11\,LMi) can be attributed to subgiants (we mark it as BB SG) and one (HD\,191588) belongs to the Galactic thick disc (we mark it as BB Thick). Among the AB stars, *37\,Com is considered to be in the He-core burning phase (cf. \citealt{Auriere2015}, \citealt{Tsvetkova2014})  and we mark it as AB~Clump. The evolution stage of our sample stars is presented in Table~\ref{tab:cccn} along with the mass, \textsuperscript{12}C/\textsuperscript{13}C, and C/N ratios.

\begin{figure}
    \includegraphics[width=\columnwidth]{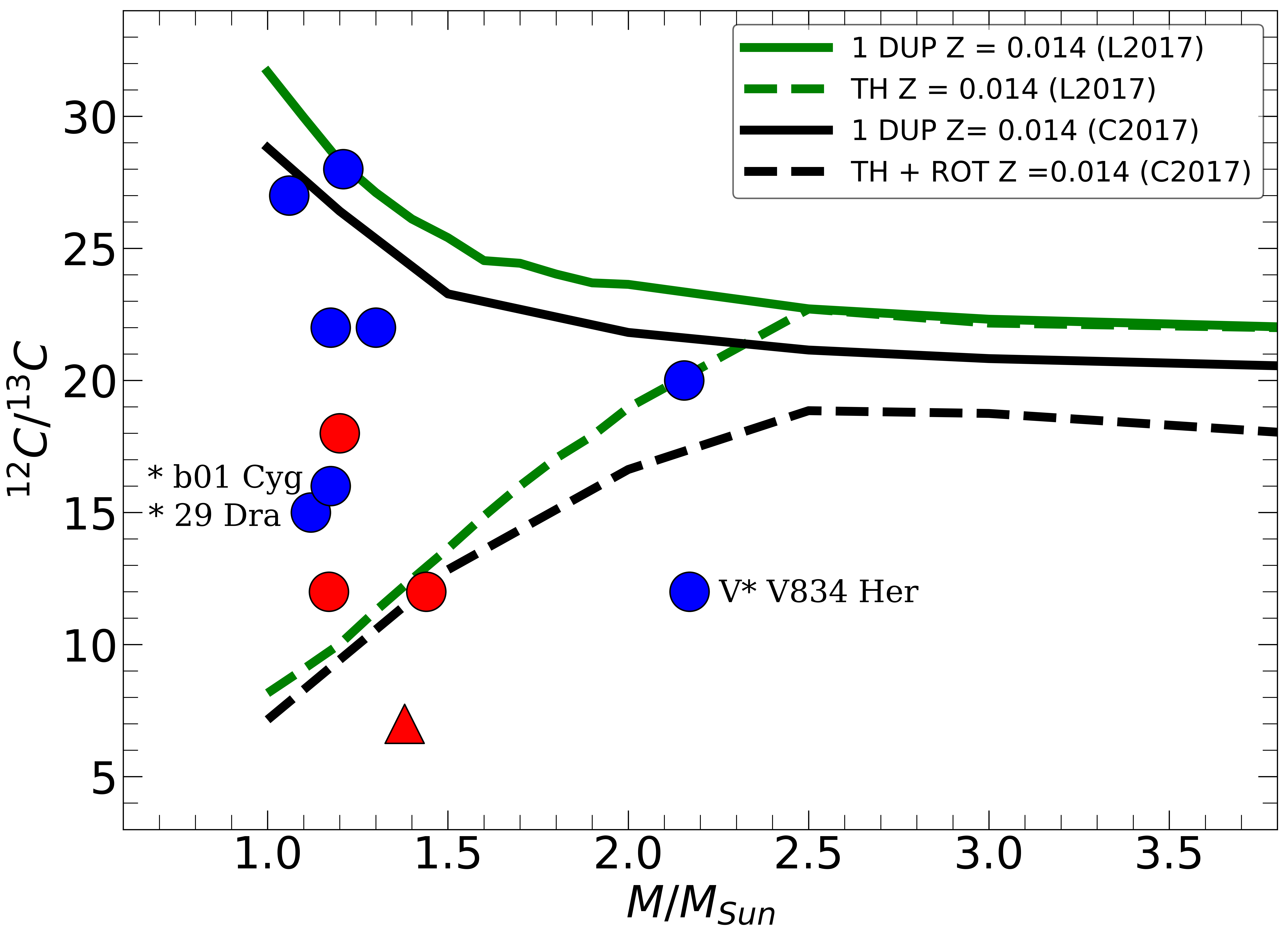}
        \includegraphics[width=\columnwidth]{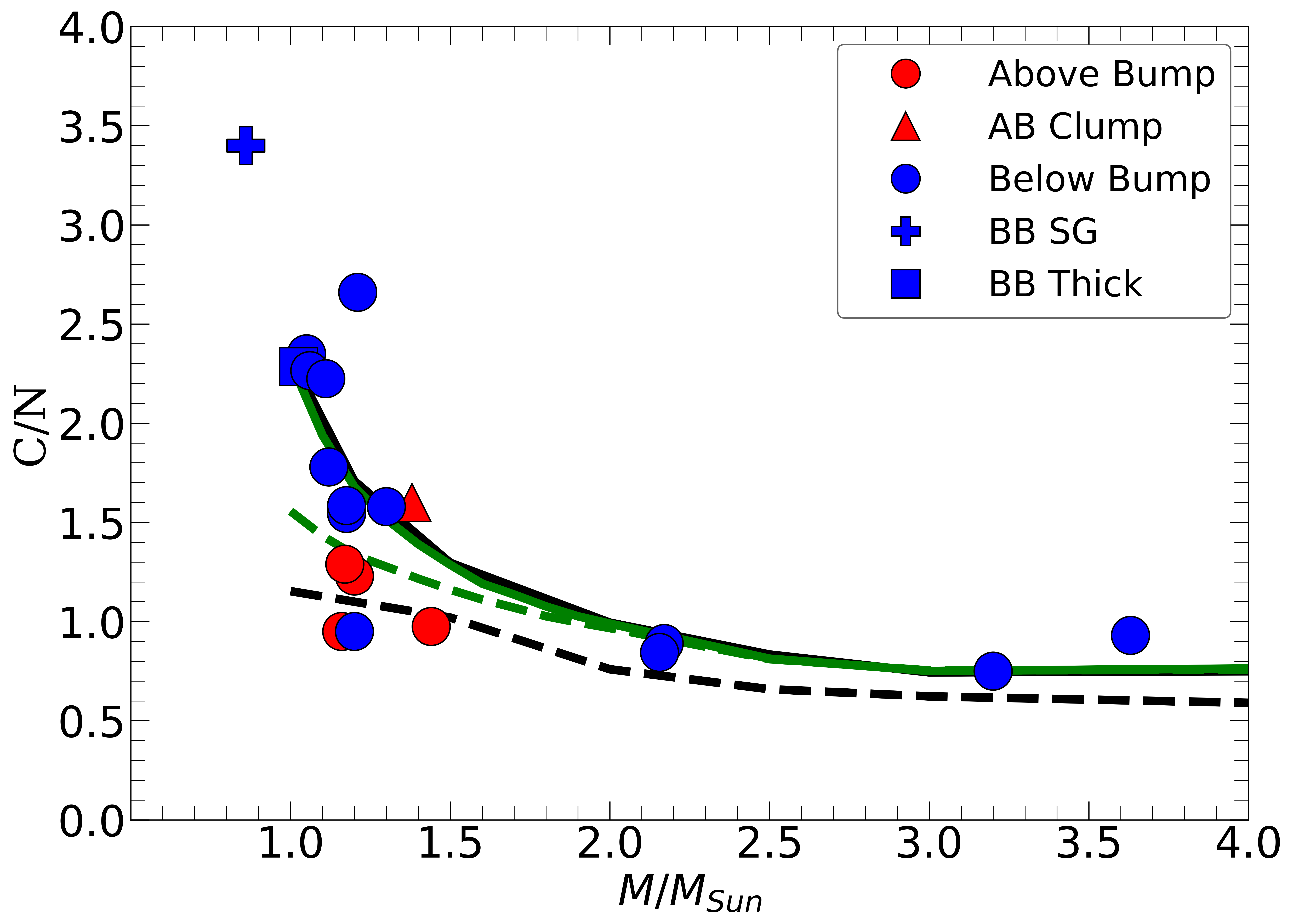}
    \caption {Ratios of \textsuperscript{12}C/\textsuperscript{13}C and C/N of the investigated stars compared with theoretical models. The solid green line represents the \textsuperscript{12}C/\textsuperscript{13}C  and  C/N ratios predicted by the model for stars at the first dredge-up (1\,DUP) and the dashed green line represents the minimal values in the model with  thermohaline-induced extra mixing (TH), both taken from \cite{Lagarde2017}. The solid black line represents the \textsuperscript{12}C/\textsuperscript{13}C  and  C/N ratios predicted by the model for stars at the first dredge-up (1\,DUP) and the dashed black line represents the minimal values in the model with thermohaline- and rotation-induced extra mixing (TH+ROT), both taken from \cite{Charbonnel2017}. Symbols are the same as in Fig.~\ref{alpha}.} 
    \label{CC}
\end{figure}

\begin{figure}
    \includegraphics[width=\columnwidth]{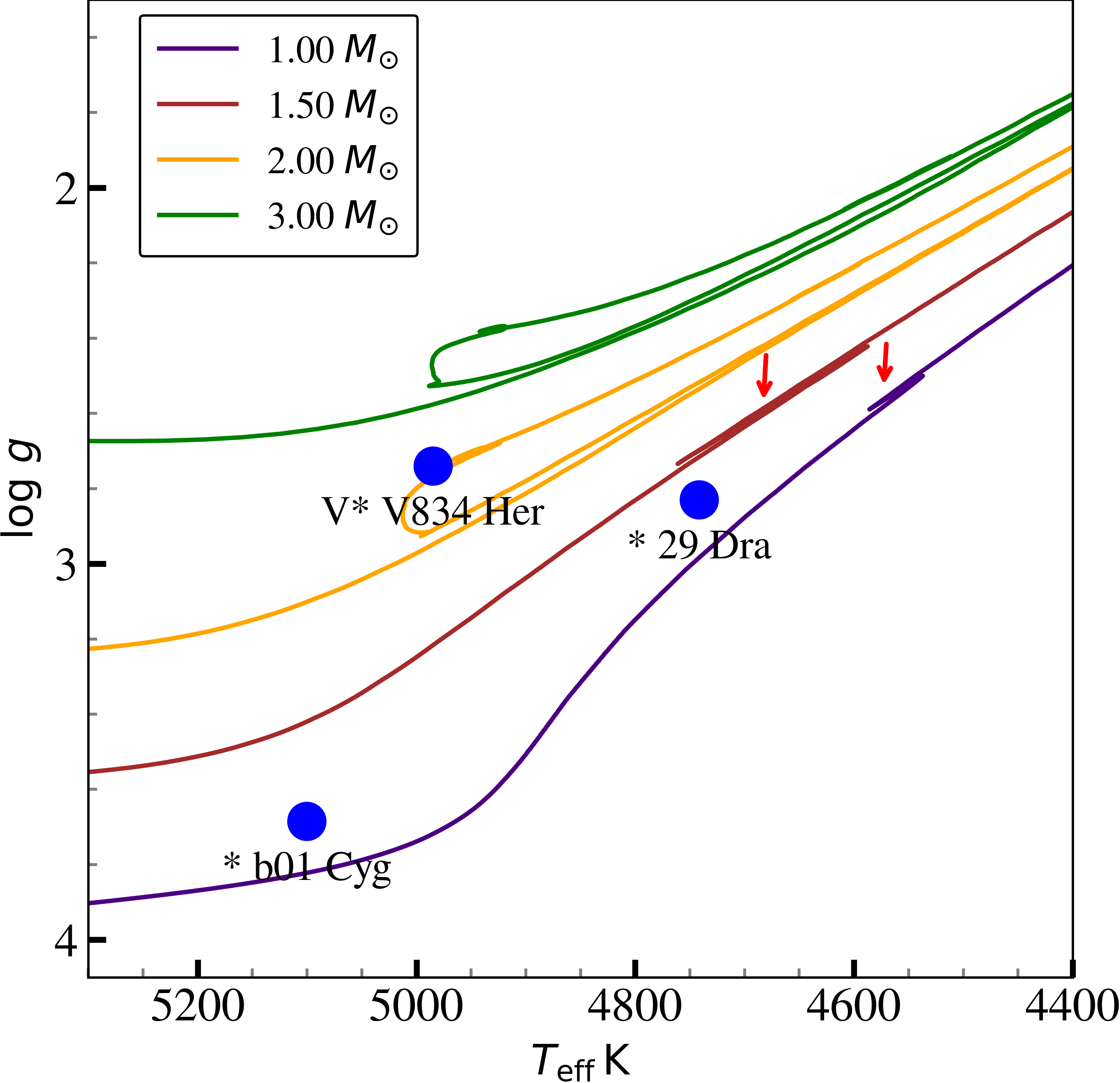}
    \caption{Stars *29\,Dra (1.1\,$M_{\odot}$), *b01\,Cyg (1.2\,$M_{\odot}$), and V*\,V834\,Her (2.2\,$M_{\odot}$) displayed in the  log\,$g$ versus $T_{\rm eff}$ diagram along with the PARSEC evolutionary tracks taken from \cite{Bressan2012} for the Solar metallicity. The red arrows show the locations of the red giant branch luminosity bumps.  
}
    \label{set}
\end{figure}

\begin{table}
\centering
\caption{Mass, $^{12}$C/$^{13}$C, C/N, and the evolutionary stage of the sample stars.}
\label{tab:cccn}
\begin{tabular}{lcccl}
\hline
\hline
Name        & Mass  & $^{12}$C/$^{13}$C  & C/N   & Evol.\\
        &  $M_{\odot}$ &     &      & stage         \\
\hline
*39 Cet         & 1.10  &  & 2.19 & BB \\ 
*39 Cet$^{\dagger}$       & 1.00  &    & 2.51 & BB \\ 
V* BF Psc     & 1.06  &  & 2.00  &  BB  \\ 
V* BF Psc$^{\dagger}$   & 1.16  &  & 2.45 & BB  \\ 
*3 Cam         & 3.20   &  & 0.75 & BB \\ 
V* CL Cam     & 1.20   &    & 0.95 & BB\\ 
V* V403 Aur    & 1.16  &   & 0.95 & AB \\ 
*ksi Leo       & 1.30   & 22  & 1.58 & BB \\ 
*11 LMi        & 0.86  &  & 3.40  & BB SG\\ 
*eps Leo        & 3.63  &  & 0.93 & BB  \\ 
*mu. Leo        & 1.20   & 18  & 1.23 & AB \\ 
*37 Com        & 1.38  & 7 & 1.60  & AB Clump\\ 
HD 145742      & 1.42  & 12  & 1.02 & AB \\ 
HD 145742 $^{\dagger}$     &  1.46 & 12  & 0.93 & AB   \\
*29 Dra         & 1.11   & 15  & 1.86 & BB \\ 
*29 Dra$^{\dagger}$       & 1.13   &     & 1.70  & BB \\ 
V* V834 Her   & 2.17  & 12  & 0.89 & BB \\ 
BD+15 3367    & 1.17  & 12  & 1.29 & AB \\ 
*62 Ser        & 1.16  & 22  & 1.55 & BB \\ 
*62 Ser$^{\dagger}$      & 1.19  & 22  & 1.54 & BB \\ 
*b01 Cyg       & 1.19  &  & 1.47 & BB \\ 
*b01 Cyg$^{\dagger}$     & 1.16  & 16  & 1.70  & BB \\ 
HD 191588      & 1.02  &  & 2.23 & BB Thick \\ 
HD 191588$^{\dagger}$    & 1.03  &  & 2.34 & BB Thick \\ 
HD 200740     & 2.17  & 20  & 0.78 & BB \\ 
HD 200740$^{\dagger}$   & 2.14  &  & 0.91 & BB \\ 
V* KX Peg     & 1.06  & 27  & 2.24 & BB \\ 
V* KX Peg$^{\dagger}$   & 1.06  &  & 2.29 & BB \\ 
HD 221639     & 1.11  &  & 2.51 & BB \\ 
HD 221639$^{\dagger}$   & 1.31  & 28  & 2.81 & BB \\ 
\hline
\end{tabular}
\tablefoot{$^{\dagger}$ Results obtained from observations with $R\sim68\,000$. Other results were obtained with $R\sim36\,000$. BB or AB $-$ stars below or above the RGB luminosity bump, respectively.  SG  $-$ subgiant star. Thick -- means that the star is attributed to the Thick disc of the Galaxy.  }
\end{table}

Figure~\ref{CC} presents a comparison of the  $^{12}$C/$^{13}$C and C/N ratios of RS\, CVn stars with several recent theoretical models: the first dredge-up (1\,DUP) and the thermohaline-induced extra mixing (TH) by \cite{Lagarde2017}, also the models featuring the (1\,DUP) and thermohaline- and rotation-induced extra mixing (TH+ROT) from \cite{Charbonnel2017} for solar metallicity stars. In the \textsuperscript{12}C/\textsuperscript{13}C versus mass plot (Fig.~\ref{CC}), we see that stars that are above the RGB luminosity bump (red circles) and the clump star with the smallest carbon isotope ratio value (red triangle) are located close to the pure thermohaline-induced model and to the thermohaline- and rotation-induced model, as expected.

The majority of stars that are below the RGB luminosity bump (blue circles) align with the first dredge-up model, as expected. However, three stars that are at the evolutionary phase below the RGB luminosity bump (see Fig.~\ref{set}) have lower  \textsuperscript{12}C/\textsuperscript{13}C values than expected for their evolutionary stage in theoretical models of the first dredge-up (cf. \citealt{Lagarde2017}, \citealt{Charbonnel2017}). Those three stars (*29\,Dra, *b01\,Cyg, and V*\,V834\,Her) exhibit \textsuperscript{12}C/\textsuperscript{13}C ratios common in stars already affected by extra mixing, which starts at the RGB luminosity bump. The carbon isotope ratios in *b01\,Cyg and *29\,Dra are about 15 and are quite close to the thermohaline-induced extra-mixing model at their mass of about 1.2~$M_\odot$. The mass of V*\,V834\,Her is larger, about 2.2~$M_\odot$. For stars of this mass, the lowest value of the carbon isotope ratio, of about 17, is predicted by the thermohaline- and rotation-induced mixing model for stars above the RGB luminosity bump. However, V*\,V834\,Her is located below the RGB luminosity bump and already shows its \textsuperscript{12}C/\textsuperscript{13}C ratio lowered to about 12. So small carbon isotope ratios could exist in stars that experience rotation-induced extra mixing and have a rotation of 250~km\,s$^{-1}$ during the main-sequence phase (\citealt{Charbonnel&Lagarde2010}), but this is not applicable to V*\,V834\,Her. The \textsuperscript{12}C/\textsuperscript{13}C ratios of stars *62\,Ser and *ksi\,Leo are also below the theoretical predictions; however, having in mind uncertainties in stellar mass determination (if their mass is under-evaluated), their carbon isotope ratio of 22 is normal. We leave these two stars for  future investigation.

For stars with undetermined carbon isotope ratios, the C/N ratios were examined to detect potential extra-mixing effects; however, no significant decrease was observed, bearing in mind that uncertainties of C/N determinations  may reach $\pm 0.4$. The alterations in   \textsuperscript{12}C/\textsuperscript{13}C ratios caused by extra mixing are more prominent in stars compared to the alterations in C/N ratios. This study demonstrates the importance of carbon isotope ratios as tools for the investigation of extra-mixing phenomena.

The alteration of \textsuperscript{12}C/\textsuperscript{13}C ratios indicates that processes of extra mixing, typically associated with the post-RGB-luminosity bump stages, may begin to influence stellar interiors earlier, particularly in low-mass stars that exhibit chromospheric activity. These stars, often characterized by strong magnetic fields and active outer layers, might undergo enhanced mixing driven by dynamic internal processes. This observation challenges traditional views on the timing of such mixing events in stellar evolution. The changes in carbon isotope ratios observed in stars below the RGB luminosity bump suggest a more intricate interplay of factors including stellar mass, magnetic activity, and rotation that influence the onset and extent of mixing processes in RS\,CVn giants. These findings underscore the complexity of stellar interior dynamics and call for a revised understanding of how and when mixing processes affect the evolution of these active stars.

Our study supports the theoretical study by \cite{Charbonnel2017} who aimed to investigate the influence of magnetic fields in G, K, and M giants. Their study was prompted by spectropolarimetric observations that  revealed localised magnetic strips in the Hertzsprung-Russell diagram coincident with the regions where the first dredge-up and core-helium burning occur. They showed that dynamo processes might occur in the stellar convective envelope at two specific moments along the evolution tracks, that is, during the first dredge-up at the base of the RGB and during central helium burning in the helium-burning phase and early- Asymptotic Giant Branch. Our study of magnetically active RS\,CVn stars, which already experienced the first dredge-up, provides observational evidence for the possible impact of magnetic fields on the chemical composition of carbon and nitrogen at this evolutionary stage.

For the evaluation of stellar activity, Ca\,\textsc{ii} H \& K lines are most often used. For most active stars, those lines show quite prominent emissions in their cores. Unfortunately, Ca\,\textsc{ii} H \& K  lines were not available in our spectra. The spectral range of the UVES spectrograph spans from 4000 to 8800~\AA\ and does not include the wavelengths around 3930 and 3960~\AA\ where these lines are located. Nevertheless, following \cite{Mallik1997} we searched our sample stars for emission in the Ca\,\textsc{ii} lines at 8498, 8542, and 8662~\AA. A prominent emission is present in the spectrum of *29 Dra (see Fig. 7 for an example). No significant emissions were visible in the Ca\,\textsc{ii} lines in the spectra of other sample stars.   

\begin{figure}
    \includegraphics[width=\columnwidth]{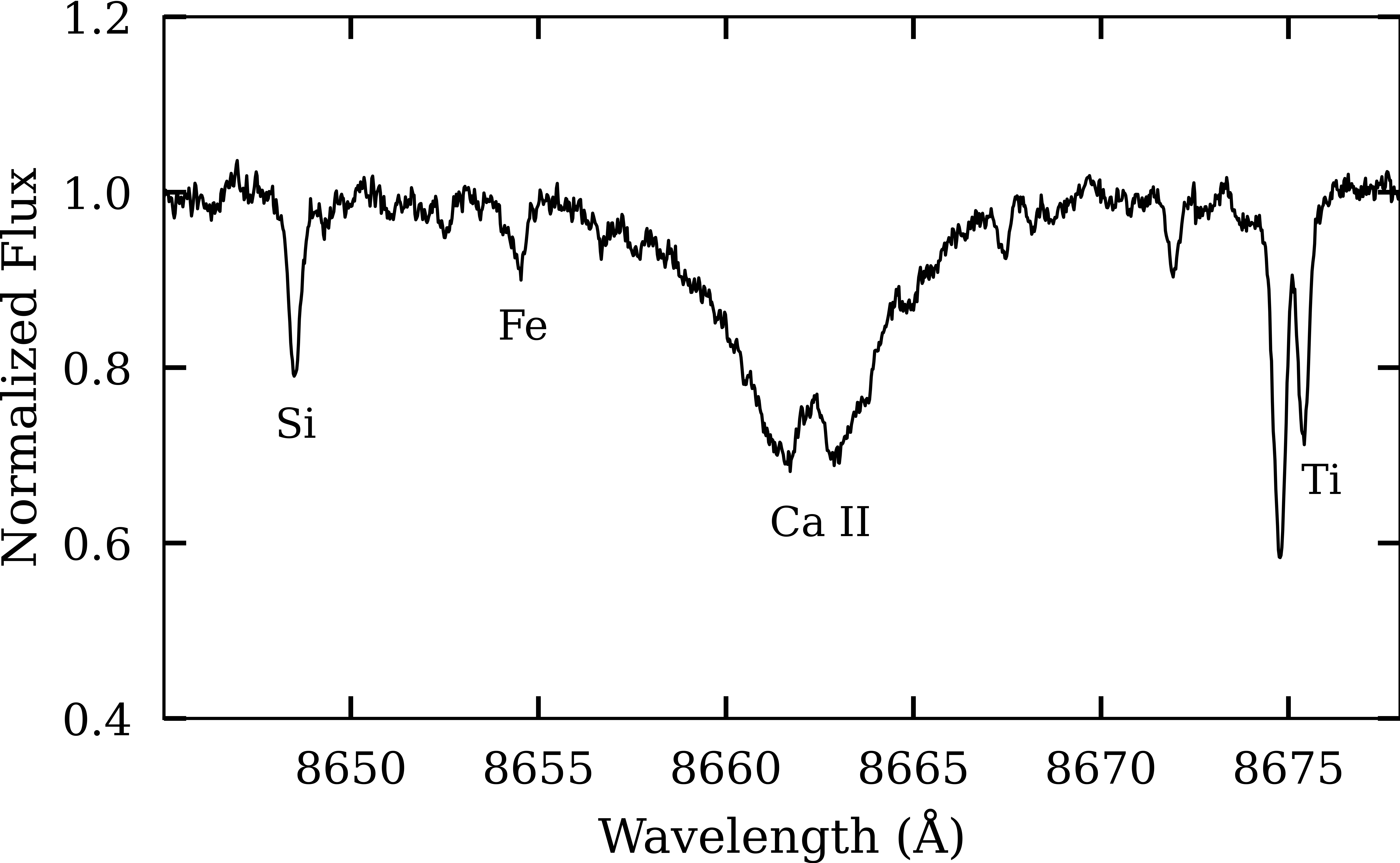}
    \caption{Example of emission in the core of Ca\,II line at 8662~\AA\ in the spectrum of *29\,Dra.}
    \label{29dra}
\end{figure}

The C/N ratios exhibit less sensitivity to mixing effects (Fig.~\ref{CC}), and the observed stars align quite well with the predictions from both the thermohaline-induced extra mixing and the first dredge-up models, which is foreseen for their corresponding evolutionary stages. The subgiant star (blue plus sign) shows the largest C/N ratio as expected. The stars *b01\,Cyg, *29\,Dra, and V*\,V834\,Her, which show lower than expected carbon isotope ratios in our study, have no visible peculiarities in their C/N ratios.  

\begin{figure*}
    \centering
    \includegraphics[width=0.95\textwidth]{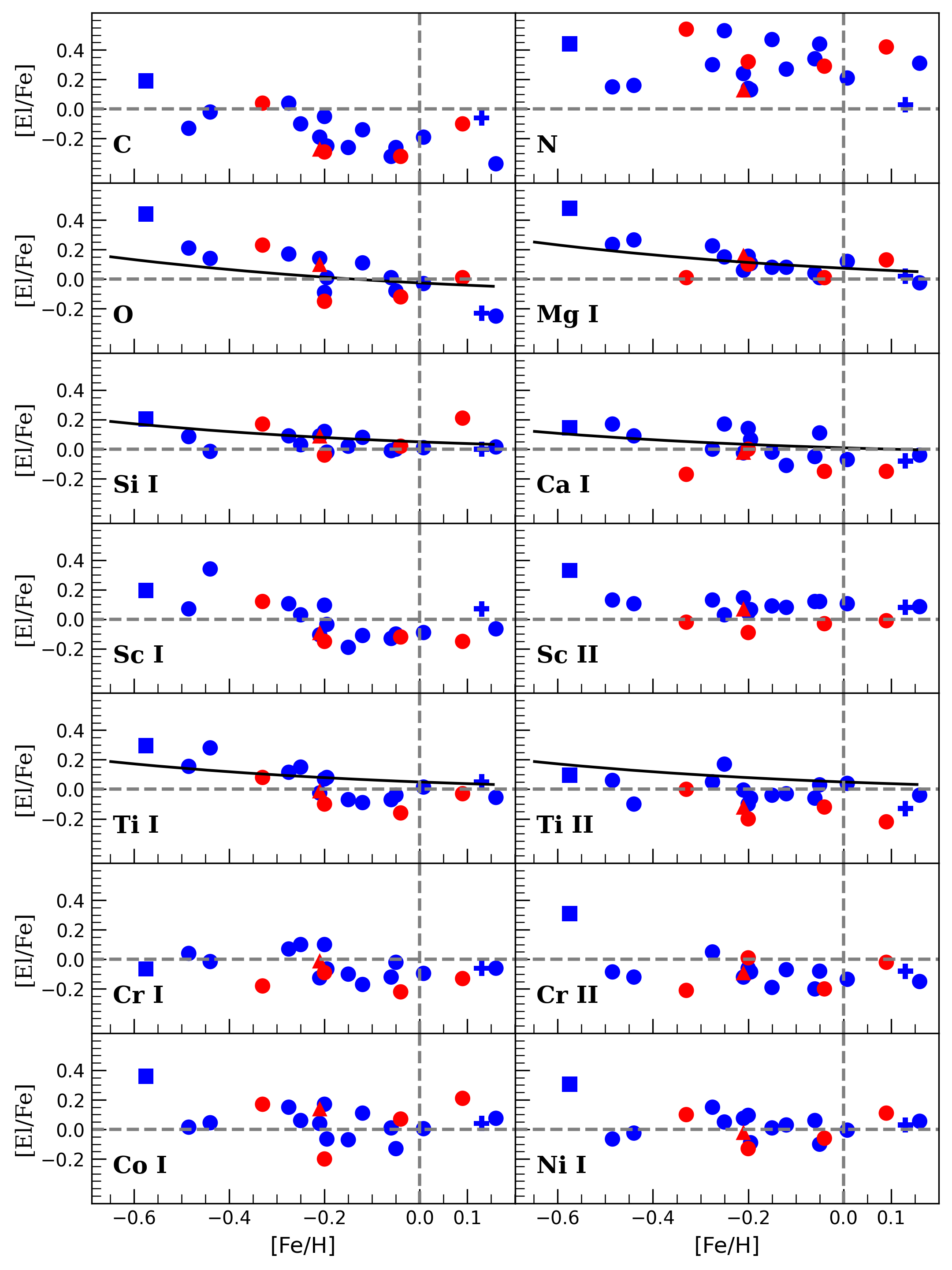}
    \caption{Ratios [El/Fe] as a function of [Fe/H]. The solid lines represent the  trends of  Galactic chemical evolution modelled by \cite{Pagel1995}. The meaning of the symbols is the same as in Fig.~\ref{alpha}. }
    \label{allele}
\end{figure*}

\begin{figure}
    \centering
    \includegraphics[width=0.95\columnwidth]{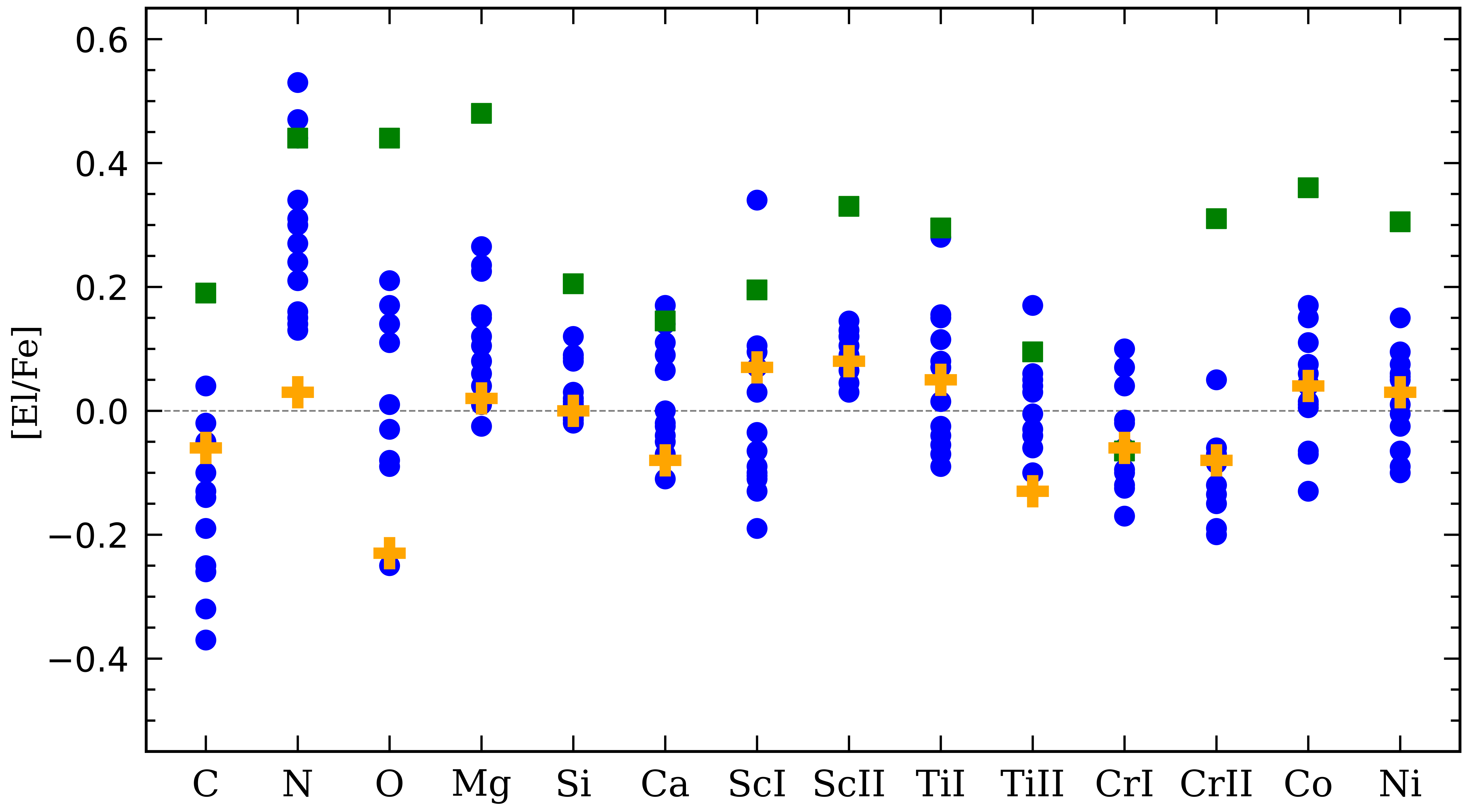}
        \includegraphics[width=1.0\columnwidth]{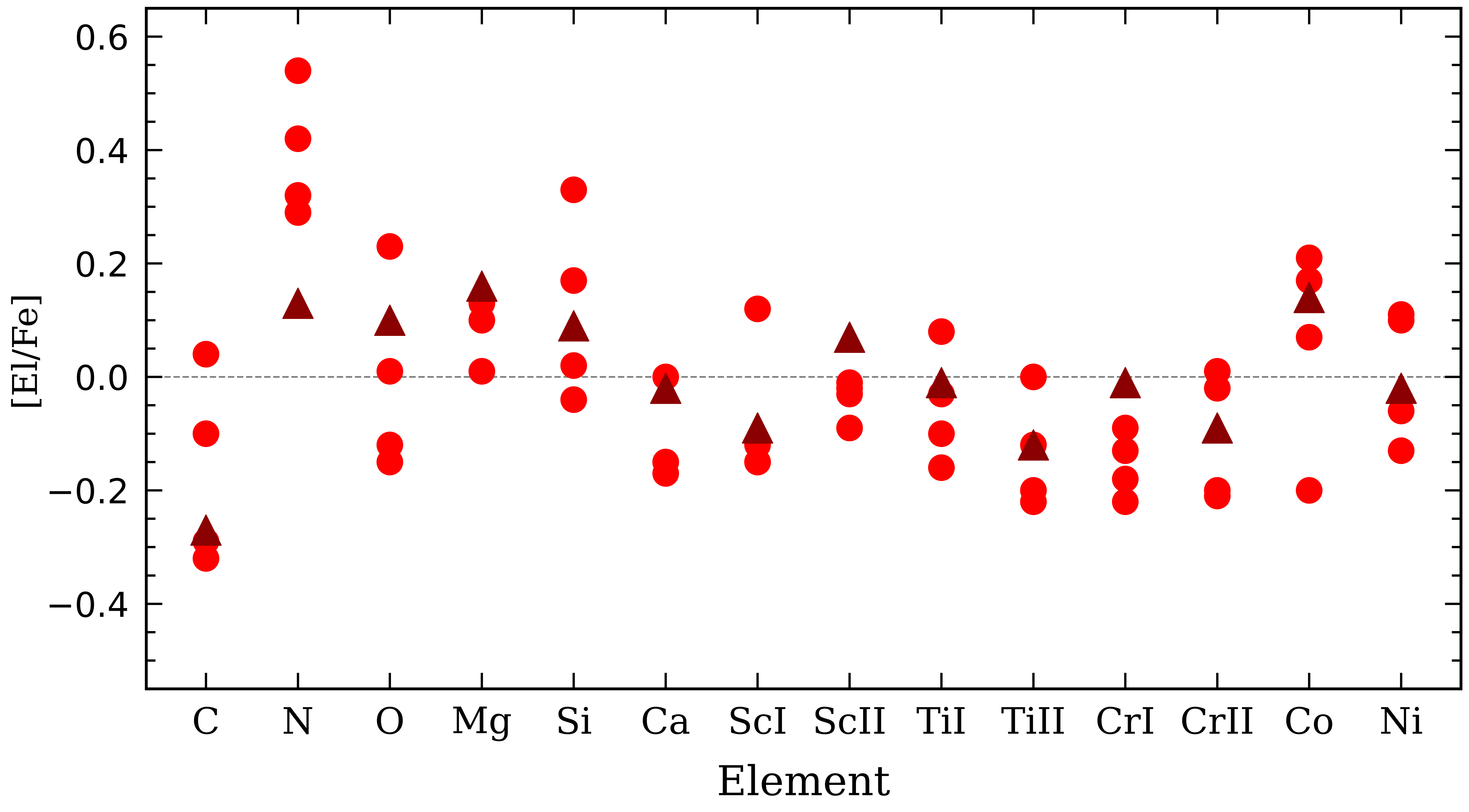}
    \caption {Ratio values of [El/Fe] for different elements in stars below the RGB luminosity bump (upper panel) and stars above the bump (lower panel). For clarity, the thick disc star is shown as the green square, the subgiant star as the orange plus sign, and the clump star as the brown triangle. }
    \label{spread16cno}
\end{figure}

\subsection{Element-to-iron abundance ratios}
\label{abun}

 The element-to-iron, [El/Fe], trends in respect to [Fe/H] are shown in Fig.~\ref{allele} together with the modelled trends of the Galactic evolution (\citealt{Pagel1995}). Despite being chromospherically active, the investigated RS\,CVn stars do not show systematic differences from the Galactic evolutionary models.  The thick disc star is enhanced in abundances of light elements as expected. Its carbon abundance is also larger in comparison to abundances in other giants located below the RGB luminosity bump. Ratios of [C/Fe] in the latter are predominantly lower than Solar. 
 Ratios of [N/Fe]  are enhanced in all the stars, except the one that is in the evolutionary stage of subgiants, which is not yet affected by the first dredge-up and extra mixing. This star has an unchanged chemical composition.  
 
In Fig.~\ref{spread16cno}, a similar spread of [El/Fe] is visible for the investigated chemical elements in stars below and above the RGB luminosity bump. As expected, the subgiant star of slightly super-solar metallicity shows close to Solar [El/Fe] ratios, including the carbon and nitrogen abundances, which reflects its evolutionary stage before significant surface convective mixing occurred. Elemental abundances of this subgiant star might closely represent the initial composition of the gas clouds from which it was formed. Other stars show a wider spread in their [El/Fe] values, especially for the mixing-sensitive elements C and N. This spread indicates varying degrees of internal mixing, which depends on the stellar mass, metallicity, evolutionary phase, and other parameters. Ratios of [El/Fe]  in the investigated clump star are similar to those of stars that are above the RGB luminosity bump. The thick-disc star, in comparison to other below-RGB luminosity bump giants, has larger abundances as expected.

\section{Summary and conclusions} 
\label{conclusion}
The research presented in this paper focused on a sample of 20 RS\,CVn stars observed in high spectral resolution with the 1.65~m Moletai Observatory telescope and the UVES spectrograph, and examined their atmospheric parameters and chemical abundances of $^{12}$C, $^{13}$C, N, O, Mg\,\textsc{i}, Si\,\textsc{i}, Ca\,\textsc{i}, Sc\,\textsc{i}, Sc\,\textsc{ii}, Ti\,\textsc{i}, Ti\,\textsc{ii}, Cr\,\textsc{i}, Cr\,\textsc{ii}, Fe\,\textsc{i}, Fe\,\textsc{ii}, Co, and Ni. The study aimed to provide insights into the internal mixing processes that act in these chromospherically active stars through the analysis of chemical signatures.

Stellar atmospheric parameters were consistently determined using the classical equivalent width method. Elemental abundances were determined using spectral synthesis for C, N, O, and Mg,\textsc{i}, and the equivalent width method was used for elements with a larger number of available lines. The observed stars, except one subgiant and one He-core burning clump star, are in evolutionary stages along the RGB. Metallicity measurements varied from $-0.6$ to 0.17~dex, with an average slightly below the Solar value.

We also calculated Galactic space velocities \textit{(U, V, W)} and orbital parameters, as well as stellar masses and ages for our sample stars. Based on the Toomre diagram, the $[\alpha/\text{Fe\,\textsc{i}}]$ versus [Fe/H] diagram, and the star's position relative to the RGB luminosity bump in the HR diagram, we attributed 15 stars of our sample to the evolutionary stage below the RGB luminosity bump and five to the stage above the bump. Among them, one star was attributed to subgiants, one to the He-core burning phase, and one to the Galactic thick disc. 

We calculated the \textsuperscript{12}C/\textsuperscript{13}C and  C/N ratios for the stars, and determined that *29\,Dra, *b01\,Cyg, and V*\,V834\,Her, which are in the evolutionary stage below the RGB luminosity bump, already show evidence of extra mixing in their lowered carbon isotope ratios as was determined in \(\lambda\)~And by \cite{Tautvaisiene2010b} and \cite{Drake2011} and in \*29\,Dra by \cite{Barisevicius2010}. With this work, we provide observational evidence that in some low-mass chromospherically active RS\,CVn stars  extra-mixing processes may start to act below the luminosity bump of the RGB, that is, earlier than in inactive giants. 

\section*{Data availability} 
\label{dataava}
Full Table \ref{table:Results} is available at the CDS via anonymous ftp to cdsarc.u-strasbg.fr (130.79.128.5) or via http://cdsweb.u-strasbg.fr/cgi-bin/qcat?J/A+A/.

\begin{acknowledgements}
We acknowledge funding from the Research Council of Lithuania (LMTLT, grant No. S-MIP-23-24). 
This work has made use of data from the European Space Agency (ESA) mission
{\it Gaia} (\url{https://www.cosmos.esa.int/gaia}), processed by the {\it Gaia} Data Processing and Analysis Consortium (DPAC,
\url{https://www.cosmos.esa.int/web/gaia/dpac/consortium}). Funding for the DPAC has been provided by national institutions, in particular, the institutions participating in the {\it Gaia} Multilateral Agreement.
We have made extensive use of the NASA ADS and SIMBAD databases.
\end{acknowledgements}

\bibliographystyle{aa}
\bibliography{ref.bib}

\onecolumn
\appendix 
\section{Machine readable table of results}

\begin{tiny}
 \begin{longtable}{llll}
 \caption{Contents of the machine-readable table available online at the CDS.}
 \label{table:Results}\\
 \hline
 \hline
 Col & Label & Units & Explanations\\
 \hline
 1      & ID                 & --          & Tycho-2 catalogue identification\\
 2   & Name               & --           & Stellar name \\
 3   & Res                & --          & Spectral resolution   \\
 4      & $T_{\rm eff}$     & K             & Effective temperature\\
 5      & $e$\_$T_{\rm eff}$  & K        & Uncertainty in effective temperature\\
 6      & Log\,$g$               & dex & Stellar surface gravity\\
 7      & $e$\_Log\,$g$            & dex & Uncertainty in stellar surface gravity\\
 8      & [Fe/H]             & dex          & Metallicity\\
 9      & $e$\_Fe\,\textsc{i}          & dex        & Uncertainty in [Fe\,\textsc{i}/H]\\
 10   & n\_Fe\,\textsc{i}             &  --         & Number of Fe\,\textsc{i} lines \\
 11  & $e$\_Fe\,\textsc{ii}            &  dex        & Uncertainty in [Fe\,\textsc{ii}/H] \\
 12  & n\_Fe\,\textsc{ii}            &   --          & Number of Fe lines \\
 13     & $V_{\rm t}$         &  km\,s$^{-1}$   & Microturbulence velocity\\
 14     & $e\_Vt$              & km\,s$^{-1}$  & Uncertainty in microturbulence velocity\\
 15     & [C/H]                  & dex              & Carbon abundance\\
 16     & $e$\_[C/H]         & dex          & Uncertainty in carbon abundance\\
 17  & n\_C                  & --           & Number of C$_2$ lines \\
 18     & [N/H]                  & dex              & Nitrogen abundance\\
 19     & $e$\_[N/H]         & dex          & Uncertainty in nitrogen abundance\\
 20  & n\_N                  & --           & Number of CN lines \\
 21     & [O/H]                  & dex              & Oxygen abundance\\
 22     & $e$\_[O/H]         & dex          & Uncertainty in oxygen abundance\\
 23  & n\_O                  & --           & Number of oxygen lines \\
 24     & [Mg/H]             & dex          & Magnesium abundance\\
 25     & $e$\_[Mg/H]        & dex          & Uncertainty in magnesium abundance\\
 26  & n\_Mg                  & --           & Number of magnesium lines \\
 27     & [Si/H]             & dex          & Silicon abundance\\
 28     & $e$\_[Si/H]        & dex          & Uncertainty in silicon abundance\\
 29  & n\_Si                  & --           & Number of silicon lines \\
 30     & [Ca/H]             & dex          & Calcium abundance\\
 31     & $e$\_[Ca/H]        & dex          & Uncertainty in calcium abundance\\
 32  & n\_Ca                  & --           & Number of calcium lines \\
 33     & [Sc\,\textsc{i}/H]                 & dex          & Scandium abundance from neutral lines\\
 34     & $e$\_[Sc\,\textsc{i}/H]            & dex          & Uncertainty in [Sc\,\textsc{i}/H] abundance\\
 35  & n\_Sc\,\textsc{i}                  & --           & Number of Sc\,\textsc{i} lines \\
  36    & [Sc\,\textsc{ii}/H]                & dex          & Scandium abundance from ionized lines\\
 37     & $e$\_[Sc\,\textsc{ii}/H]           & dex          & Uncertainty in [Sc\,\textsc{ii}/H] abundance\\
 38  & n\_Sc\,\textsc{ii}                  & --           & Number of Sc\,\textsc{ii} lines \\
  39    & [Ti\,\textsc{i}/H]                 & dex          & Titanium abundance from neutral lines\\
 40     & $e$\_[Ti\,\textsc{i}/H]            & dex          & Uncertainty in [Ti\,\textsc{i}/H] abundance\\
 41  & n\_Ti\,\textsc{i}                  & --           & Number of Ti\,\textsc{i} lines \\
 42     & [Ti\,\textsc{ii}/H]                & dex          & Titanium abundance from ionized lines\\
 43     & $e$\_[Ti\,\textsc{ii}/H]           & dex          & Uncertainty in [Ti\,\textsc{ii}/H] abundance\\
 44  & n\_Ti\,\textsc{ii}                  & -- & Number of Ti\,\textsc{ii} lines \\
  45    & [Cr\,\textsc{i}/H]                 & dex          & Chromium abundance from neutral lines\\
 46     & $e$\_[Cr\,\textsc{i}/H]            & dex          & Uncertainty in [Cr\,\textsc{i}/H] abundance\\
 47  & n\_Cr\,\textsc{i}                  & --           & Number of Cr\,\textsc{i} lines \\
 48     & [Cr\,\textsc{ii}/H]                & dex          & Chromium abundance from ionized lines\\
 49     & $e$\_[Cr\,\textsc{ii}/H]           & dex          & Uncertainty in [Cr\,\textsc{ii}/H] abundance\\
  50  & n\_Cr\,\textsc{ii}                  & --           & Number of Cr\,\textsc{ii} lines \\ 
  51    & [Co/H]             & dex          & Cobalt abundance\\
 52     & $e$\_[Co/H]        & dex          & Uncertainty in cobalt abundance\\
 53  & n\_Co                  & --           & Number of cobalt lines \\
54      & [Ni/H]             & dex          & Nickel abundance\\
 55     & $e$\_[Ni/H]        & dex          & Uncertainty in nickel abundance\\
 56  & n\_Ni                  & --           & Number of nickel lines \\     
 57   & $^{12}$C/$^{13}$C & --           & Carbon isotope ratio \\
 58   & $e$\_$^{12}$C/$^{13}$C & --        & Uncertainty in carbon isotope ratio \\
 59     & C/N               & --           & Carbon-to-nitrogen abundance ratio \\
  60    & Mass              & $M_\odot$    & Stellar mass \\
  61    & $e$\_Mass         & $M_\odot$    & Uncertainty of mass \\
  62    & Age               & Gyr          & Stellar age \\
  63    & $e$\_Age          & Gyr          & Uncertainty of age \\
  64    & $R_{\rm mean}$         & kpc          & Mean galactocentric distance \\
  65    & $e$\_$R_{\rm mean}$  &  kpc      & Uncertainty of mean galactocentric distance \\
 \hline
 \end{longtable}
 \centering
\end{tiny}

\end{document}